\documentclass[%
 amsmath,amssymb,
prf,
floatfix,
longbibliography,
notitlepage,
onecolumn
]{revtex4-1}

\usepackage{physicstests}
\usepackage{dcolumn}% Align table columns on decimal point
\usepackage{bm}% bold math
\usepackage{caption}
\usepackage{subcaption}
\usepackage{cancel}
\usepackage{booktabs}

\newcommand{\fd}{\mbox{$\phi_{\epsilon}$}}
\newcommand{\Gd}{\mbox{$G_{\epsilon}$}}
\newcommand{\Bd}{\mbox{$B_{\epsilon}$}}
\newcommand{\bx}{\mbox{\bf x}}

\newcommand{\PE}{$\phi_\epsilon$}
\newcommand{\SE}{$\psi_\epsilon$}

\newcommand{\by}{\mbox{\bf y}}

\newcommand{\ff}{\mbox{\bf f}}
\newcommand{\bR}{\mbox{\bf R}}

\begin{document}

\title{Using theory and experiments of spheres moving near boundaries to optimize the method of images for regularized Stokeslets}

\author{Hoa Nguyen}
\email{hnguyen5@trinity.edu}

\author{Amelia Gibbs}
 \email{tgibbs@trinity.edu}

 \author{Frank Healy}
\email{fhealy@trinity.edu}

 \author{Orrin Shindell}
 \email{oshindel@trinity.edu}
 \affiliation{%
  Trinity University, San Antonio, Texas
 }%

\author{Ricardo Cortez}
\affiliation{%
 Tulane University, New Orleans, Louisiana
}%
\email{rcortez@tulane.edu}

\author{Kathleen M. Brown}
 \email{kate.brown@centre.edu}
 
 \author{Jonathan McCoy}
\email{jonathan.mccoy@centre.edu}

\author{Bruce Rodenborn}
\email{bruce.rodenborn@centre.edu}
  \affiliation{Centre College, Danville, Kentucky}

%\collaboration{Trinity--Centre Collaboration}%\noaffiliation


\begin{abstract}
The general system of images for regularized Stokeslets (GSIRS) developed by Cortez and Varela (2015) is used extensively to model Stokes flow phenomena such as microorganisms swimming near a boundary. Our collaborative team uses  dynamically similar scaled macroscopic experiments to test theories for forces and torques on spheres moving near a boundary and use these data and the method of regularized Stokeslets (MRS) created by Cortez et al. (2005) to calibrate the GSIRS. We find excellent agreement between theory and experiments, which provides the first experimental validation of exact series solutions for spheres moving near an infinite plane boundary. We test two surface discretization methods commonly used in the literature: the six-patch method and the spherical centroidal Voronoi tessellation (SCVT) method. Our data show that the SCVT method provides the most accurate results when the motional symmetry is broken by the presence of a boundary. We use theory and the MRS to find optimal values for the regularization parameter in free space for a given surface discretization and show that the optimal regularization parameter values can be fit with simple formulae when using the SCVT method.  We also present a regularization function with higher order accuracy when compared with the regularization function previously introduced by Cortez et al. (2005). The simulated force and torque values compare very well with experiments and theory for a wide range of boundary distances. However, we find that for a fixed discretization of the sphere, the simulations lose accuracy when the gap between the edge of the sphere and the wall is smaller than the average distance between grid points in the SCVT discretization method. We also show an alternative method to calibrate the GSIRS to simulate sphere motion arbitrarily close to the boundary. Our computational parameters and methods along with our MATLAB and PYTHON implementations of the series solution of Lee and Leal (1980) provide researchers with important resources to optimize the GSIRS and other numerical methods, so that they can efficiently and accurately simulate spheres moving near a boundary.

\end{abstract}

\maketitle

\section*{\label{sec:intro}Background}
The method of regularized Stokeslets \cite{cortez2015general} and the general system of images for regularized Stokeslets \cite{cortez2005method} have been used extensively to model swimming microorganisms.  The work of Shindell et al. (2021) \cite{shindell2021bacmotilitysurface} showed how dynamically similar macroscopic experiments and theory provide a principled method to calibrate the MRS and GSIRS. They considered a cylinder and helix moving near a boundary and showed that the optimal regularization parameter depended on both the surface discretization and geometry. 
Accurate forces and torques on a model bacterium could then be calculated and swimming performance measures such as the Purcell efficiency \cite{purcell1997efficiency}, energy per distance \cite{li2006low, li2017flagellar}, and metabolic energy cost (energy per distance per body mass) \cite{shindell2021bacmotilitysurface} could be computed. Thus, accurately calibrated models allow determination of important structure function relationships when microorganisms are moving in their environment.

We use similar techniques to calibrate the GSIRS for spherical geometries, which can be used to model \textit{Rhodobacter sphaeroides} and \textit{Enterococcus saccharolyticus} and other microorganisms with round bodies \cite{Manson_et_al_1980, Imhoff_et_al_1984, Turner_et_al_2016}. The theoretical analysis of forces and torques on spheres moving near a boundary in Stokes flow has a long history beginning with the work of Jeffrey in 1915 \cite{Jeffrey_1915} who first found a series solution to the Stokes equations to calculate the torque on a sphere rotating with its axis perpendicular to an infinite plane. Subsequent work found the drag on a sphere moving parallel to and perpendicular to a boundary \cite{Brenner_1961, oneill1964spherenearwall}, and the torque on a sphere with its rotation axis oriented parallel to the boundary \cite{Dean_ONeill_1963_rotating_sphere}. The linearity of the Stokes solutions allows any arbitrary motion of a sphere relative to a boundary to be decomposed into a sum of these four fundamental motions \cite{Lee_Leal_1980sphere_motion}. 

Work by Lee and Leal (1980) \cite{Lee_Leal_1980sphere_motion} unified these various series solutions of sphere motion near a boundary into one linear system of equations that calculates the force and torque for motion near surfaces with varying properties. However, the predicted drag and torque have not been thoroughly verified with experiments \cite{Lee_Leal_1980sphere_motion}. Previous experimental work to verify theory has generally focused on either sedimentation experiments, such as those by Malysa and van de Ven (1986)  \cite{Maysa_vandeVen_1986experiments}, or on microscopic experiments \cite{Dufresne_et_al_2000, Pitois_et_al_2009, Wang_et_al_2009, Wu_et_al_2021}. There have been fewer experimental tests of torque on a sphere such as Maxworthy (1965) \cite{maxworthy_1965} and Kunesh et al. (1985) \cite{Kunesh_et_al_1985}, but they studied a sphere in a rotating tank rather than a rotating sphere in a stationary tank. Our dynamically similar experiments provide much more complete and higher precision tests of theory than has been previously reported.  We also provide numerical implementations of the Lee and Leal theory in user-friendly MATLAB and PYTHON GUI codes in the Supplemental Information. 

We use the theory and experiments to precisely calibrate our GSIRS models, which include two regularization functions: the commonly used \PE\ developed in Cortez et al. (2005) \cite{cortez2005method} and a new regularization function, \SE. We also use two methods of surface discretization: spherical centroidal Voronoi tessellation (SCVT), as developed by Du et al. (2003) \cite{Du2003}, and the six-patch method \cite{Sadourny_1972_6patch}. We find optimal regularization parameters by minimizing the percent error between simulations and theory in free space and show that \SE\ improves the order of accuracy in GSIRS. We also show that the SCVT outperforms the six-patch discretization scheme because it is less sensitive to the orientation of the discretization to the boundary. However, we find that even the optimized GSIRS loses accuracy when the gap between the sphere and the boundary is smaller than the average distance between Stokeslets on the sphere. 
This result provides researchers a new general rule of thumb: the GSIRS may not be accurate within this distance and higher resolution is needed. Our library of experimental measurements and numerical implementations of theory are also reference values to aid researchers in precisely calibrating other numerical methods.

This article is organized as follows: Section \ref{sec:ex_methods} describes our dynamically similar experimental techniques to obtain the forces and torques for the four fundamental motions including removing finite size effects using theory. Section \ref{sec:sims} describes our discretization and regularization schemes including the regularization function \SE\ along with how we optimize the model using theory. Section \ref{sec:results} compares the experiments and optimized simulations to theory and discuss how to increase accuracy very near the boundary. Section \ref{sec:discussion} summarizes the performance of our calibrated model and our new empirical rule for establishing the minimum discretization needed for accuracy near a boundary. Appendices \ref{app:num_sims} and \ref{app:Lee_Leal} provide additional details about the simulations and our implementation of the theory.

\section{\label{sec:ex_methods} Experimental Methods}

We used dynamically similar macroscopic tank experiments to test the four fundamental motions of a sphere near an infinite plane: perpendicular translation, parallel translation, perpendicular axis rotation, and parallel axis rotation \cite{Lee_Leal_1980sphere_motion}. Experiments were performed in a 225-liter tank  (0.61 m $\times$ 0.61 m $\times$ 0.61 m) filled with approximately 180 liters of silicone oil  (Clearco\textsuperscript{\textregistered} PSF-60,000cSt Polydimethylsiloxane). The fluid has a density of 970 kg/m$^3$ and dynamic viscosity of $\mu = 51.78\pm 0.32\rm \, Pa\cdot s$ at 25\deg C. The viscosity was measured using a Thermo Scientific rheometer (Haake Viscotester IQ). The fluid temperature was measured each day of experimental trials and the manufacturers temperature coefficient for the fluid  ($1.00\times 10^{-6}$ kg/(m$\cdot$s)/$^\circ$C) was used to adjust the viscosity value. The high viscosity fluid ($>10^4$ times the viscosity of water) ensures that the Reynolds number  was much less than unity in all experiments ($Re\approx 10^{-3}$) so that the incompressible Stokes equations \eqref{eq:Stokes} were valid for the typical length and speed scales in the experiments. 

We calculated the predicted contribution of all six boundaries in the tank (five solid and one free surface) and then isolated the effects of one boundary  for easy comparison with theory and simulations by subtracting any significant effects from the other five boundaries.

\begin{figure}
\begin{tabular}{ll}
   (a)& (b)\\
    \includegraphics[width=0.45\columnwidth]{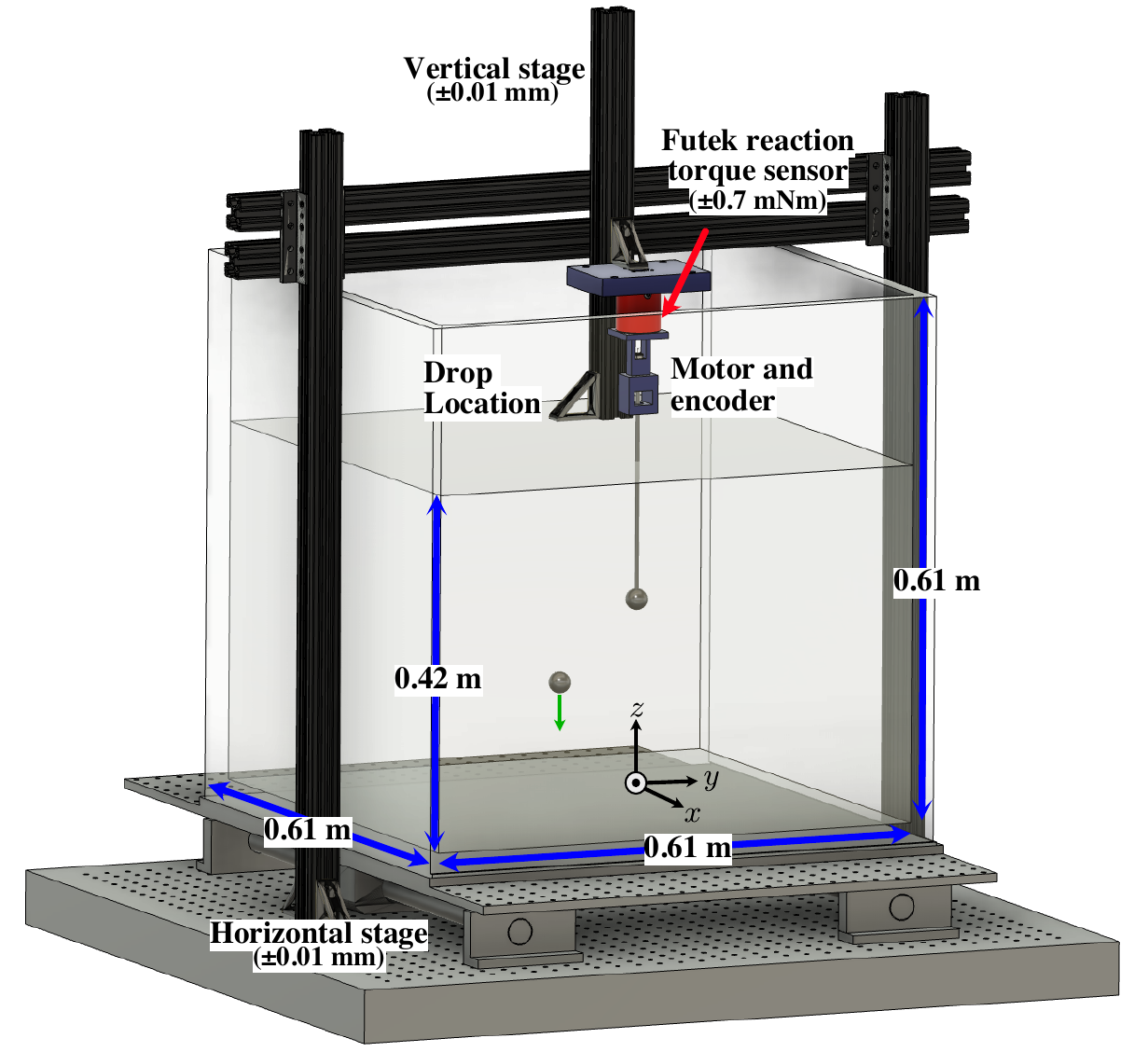} & \hspace{0.3cm}   \includegraphics[width=0.4\columnwidth]{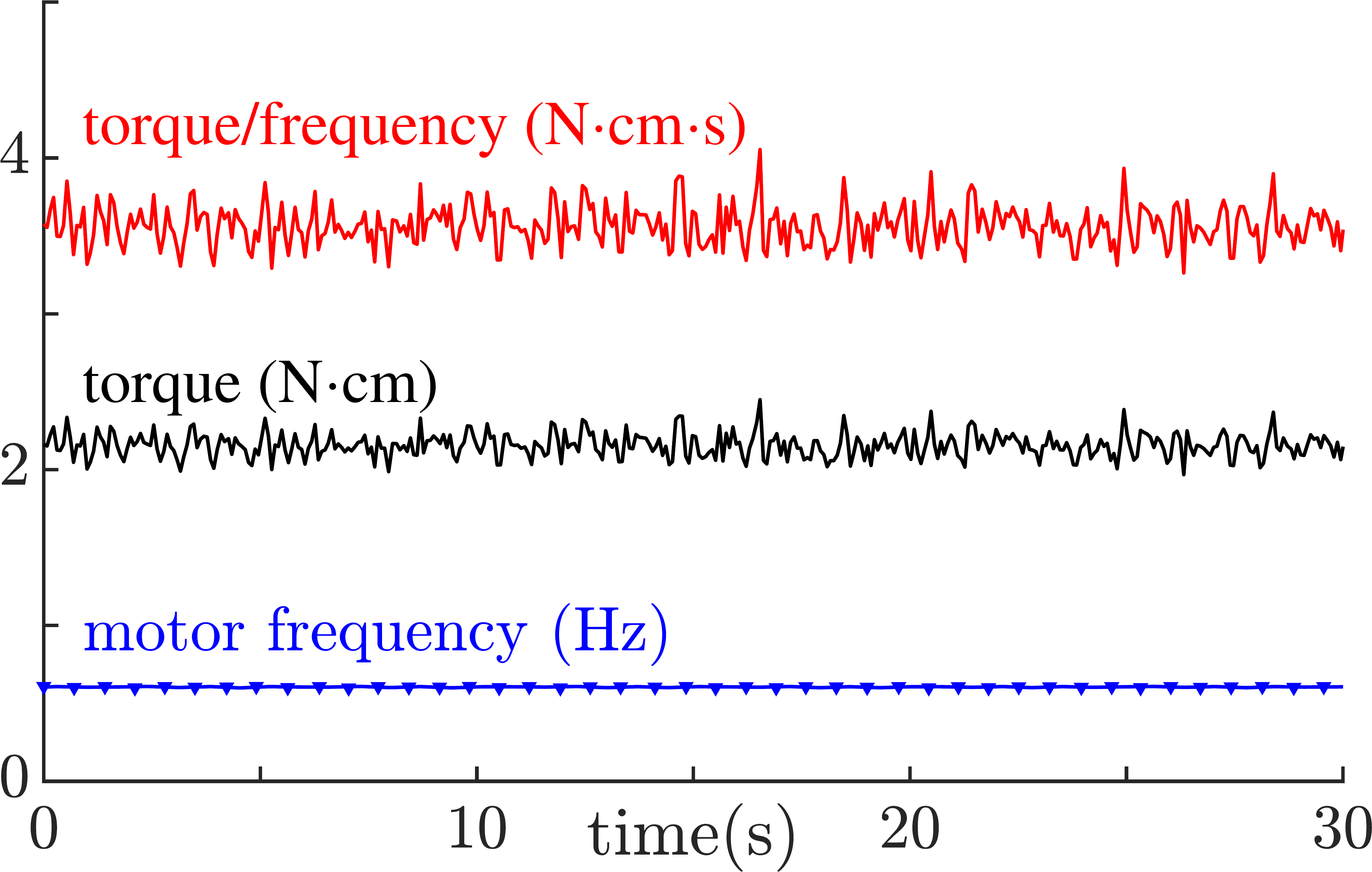}
    \\
    \end{tabular}
    \caption{Experimental setup: (a) Schematic of the tank experiments. For sedimentation experiments, spheres were held in place using a neodymium magnet at the drop location. The 80-20 extruded aluminum was used to locate the sphere relative to the front vertical wall, which was moved using the horizontal stage for parallel translation experiments. A camera and tracking software measured the speed of the sedimenting sphere. For torque experiments, vertical and horizontal stages positioned the sphere either near the bottom boundary for perpendicular axis measurements or near the front vertical boundary for parallel axis measurements. Torque was measured with a FUTEK torque sensor and the motor and magnetic encoder were housed inside a 3D-printed structure that was mounted to the active side of  sensor. (b) Example showing frequency data (solid blue symbols) read with an Agilent counter at about 1.5 samples per second, and torque data (solid black curve) were read using a DAQ at about 10 samples per second. Torque per frequency (solid red curve) was calculated by interpolating frequency data to the time where torque was measured so that dimensionless torque could be calculated. 
 }
    \label{fig:expt_setup}
\end{figure}

\subsection {Perpendicular and parallel translation} \label{sec:ex_translation}
Fluid drag was measured by sedimenting spheres with radius $R=12.70\pm 0.03$ mm % under the influence of gravity 
either perpendicular towards a boundary or parallel to a boundary. The boundary was positioned relative to the spheres using a stage, and the spheres released below the free surface using a magnet, see Figure \ref{fig:expt_setup}. We recorded motion tracking videos of falling spheres at approximately four FPS using a Point Grey GS3-U3-41C6M-C camera and a Fujinon 25 mm focal length lens positioned to minimize spherical aberration where the tracking required the most precision.  

The fluid drag was calculated as the difference between the buoyant and gravitational forces. The uncertainty in the gravitational and buoyant forces were small and considered negligible. The fluid drag in the $z-$dir, $F_\perp$ and $F_{\parallel}$, were made dimensionless by dividing by ($\mu Rv$), where $\mu$ is dynamic viscosity of the fluid, $R$ is the radius of the sphere, and $v$ is the instantaneous speed of the sphere. The speed was calculated using a central difference between the tracked positions divided by the time between frames. 

The Lee and Leal theory predicts an additional drag of about 10\% from the other five boundaries (four vertical and the upper free surface) in the location where drag measurements were made, which was subtracted from the data. We present three trials in Figure. \ref{fig:drag_torque}(a) showing very little scatter in the perpendicular drag data, so no  uncertainty was assigned for these data. When measuring the parallel drag, we tracked the spheres for $\Delta z \approx 70$ mm after they were about $ 0.10$ m below the surface, where the additional drag from the upper and lower boundaries was approximately constant. The uncertainty in the speed was typically $ \Delta v \approx 0.3$ mm/s based on uncertainty in position of the sphere, and the propagated error plotted as vertical error bars.

\subsection{Perpendicular and parallel rotation}\label{sec:ex_rotation}
Torque measurements used a Bal-Tec\textsuperscript{\textregistered} ball gauge: a stainless steel sphere of radius $R=12.70\pm 0.01$ mm with a precision hole drilled along a diameter. We removed the manufacturer's supplied shaft and attached a  carbon steel rod with radius $r = 4.763\pm 0.013$ mm and length $L = 304.8\pm 0.1$mm. The rod was attached to a DC motor with a magnetic encoder (Pololu 4846), which was mounted inside of a 3D-printed enclosure. The shaft passed through a sleeve bearing, which reduced the sphere's precession to $< 0.25$mm.  

We measured torque and rotation rate for 60-80 rotation periods  in both the CW and CCW rotation directions at each boundary location, with the difference in the two values used as the uncertainty. The DC motor was under computer control (ARDUINO MEGA 2560 with an ADAFRUIT Motor Shield v.2) and the motor frequency read using a counter (Agilent 53132A) that removed high frequency noise. A  FUTEK TFF400, 10 in-oz, Reaction Torque Sensor measured the torque, and a FUTEK Amplifier Model IAA100 amplified the signal, which was fed into a National Instruments (NI USB-6211) DAQ and recorded using MATLAB software. The frequency and torque data were read at different rates, so we interpolated frequency to the same time points that the torque was read, see Figure \ref{fig:expt_setup}(b). This interpolation was necessary so that the torque could be expressed as a dimensionless quantity for comparison to theory and simulations.

For perpendicular axis torque measurements, the rod+sphere assembly was in the middle of the tank, as shown in Figure \ref{fig:expt_setup}. The distance to the bottom boundary was adjusted using a computer-controlled vertical stage with a resolution $\pm 0.01$mm. For parallel axis measurements, the sphere was at a  depth of 0.21 and the distance to the front boundary was adjusted using computer-controlled horizontal stage with  resolution $\pm 0.01$mm. We used the theory of Jeffrey and Onishi \cite{JeffreyOnishi1981cyltorque, shindell2021bacmotilitysurface} to calculate the torque on the rod and subtracted it from the total to isolate the torque of the sphere.  Lee and Leal theory predicts negligible contributions to the sphere's torque from finite size effects($<0.1\%$), which were ignored. Torque data were made dimensionless using $(\mu \Omega R^3 )$, where $\mu$ is the dynamic viscosity, $\Omega$ is the angular speed of the rotating sphere, and $R$ is the sphere's radius. Plots of  dimensionless torque versus scaled boundary distance are shown in Figure \ref{fig:drag_torque}(c-d) along with the theory and simulations.

\section{Numerical methods}\label{sec:sims}

Our primary goal is to develop an optimized method of modeling spheres moving near a boundary using experiments and the Lee and Leal theory \cite{Lee_Leal_1980sphere_motion} to provide reference values for the optimization. We sought to minimize the two primary sources of error in any fluid simulation of a solid object: those from approximations in the method and those from discretizing the object.
%We optimize the simulations with respect two sources of error inherent in such simulations: approximations in the method and surface discretization of the solid object. %We first introduce a combination of a high-order regularization function and a surface discretization with uniform point distribution to improve the accuracy in the model and the performance of the GSIRS \cite{cortez2015general}. Then we find optimal values for the regularization parameter in free space and near a boundary. We investigate two discretization methods commonly used in the literature -- the six-patch method and the spherical centroidal Voronoi tessellation (SCVT) method to understand how the percent error relative to theory is affected by the number of grid points and the orientation of the discretized points relative to the boundary for each type of discretization. We then simulate sphere motion near a boundary for a large range of boundary distances and compare the results with experiments and theory.
We first used the MRS developed by Cortez et al. (2005) \cite{cortez2005method} to find optimal values for the regularization parameter, $\epsilon$, for a given discretization in free space and then used the GSIRS to simulate the four fundamental motions of a sphere near a surface for comparison with the theory and experiments.

The MRS uses a regularization function $\omega_\epsilon(r)$ to approximate singular point forces arranged on a surface that represents a solid object in Stokes flow. Replacing the singular forcing with a regularization function creates a linear system of equations that can be solved without using specialized methods to accommodate the singularities. The GSIRS extends the MRS to include a solid boundary by imposing a counter Stokeslet, a potential dipole, a Stokeslet doublet, and two rotlets at the image point of each discrete point $\mathbf{x}_k$. The image system thereby imposes a no-slip condition by canceling the fluid velocity at the boundary, as shown in \cite{cortez2015general}. 

The GSIRS simulations solve the incompressible Stokes equations with external forcing: 
\begin{align} \label{eq:Stokes}
\mu  \triangle \mathbf{u(\mathbf{x})} - \nabla p(\mathbf{x}) & = -\mathbf{F(\mathbf{x})} \text{ and}\\ 
\nabla \cdot \mathbf{u(\mathbf{x})} & = 0, \nonumber 
\end{align}

where $\mathbf{u}$ is the fluid velocity, $p$ is the fluid pressure, and $\mu$ is the dynamic viscosity. The vector $\mathbf{F}$ is the sum of the force density applied at discrete points $\mathbf{x}_k$, $k = 1,...,N$ on the sphere model, i.e., $\mathbf{F}(\mathbf{x}) = \displaystyle \sum_{k = 1}^N \mathbf{f}_k \omega_\epsilon(\mathbf{x} - \mathbf{x}_k)$ where $\mathbf{f}_k$ is a point force at $\mathbf{x}_k$ and $\omega_\epsilon(\mathbf{x})$ is a regularization function having an integral equal to unity over $\bR^3$. 

\begin{figure}[ht]
    \centering
    \includegraphics[width=1.0\textwidth]{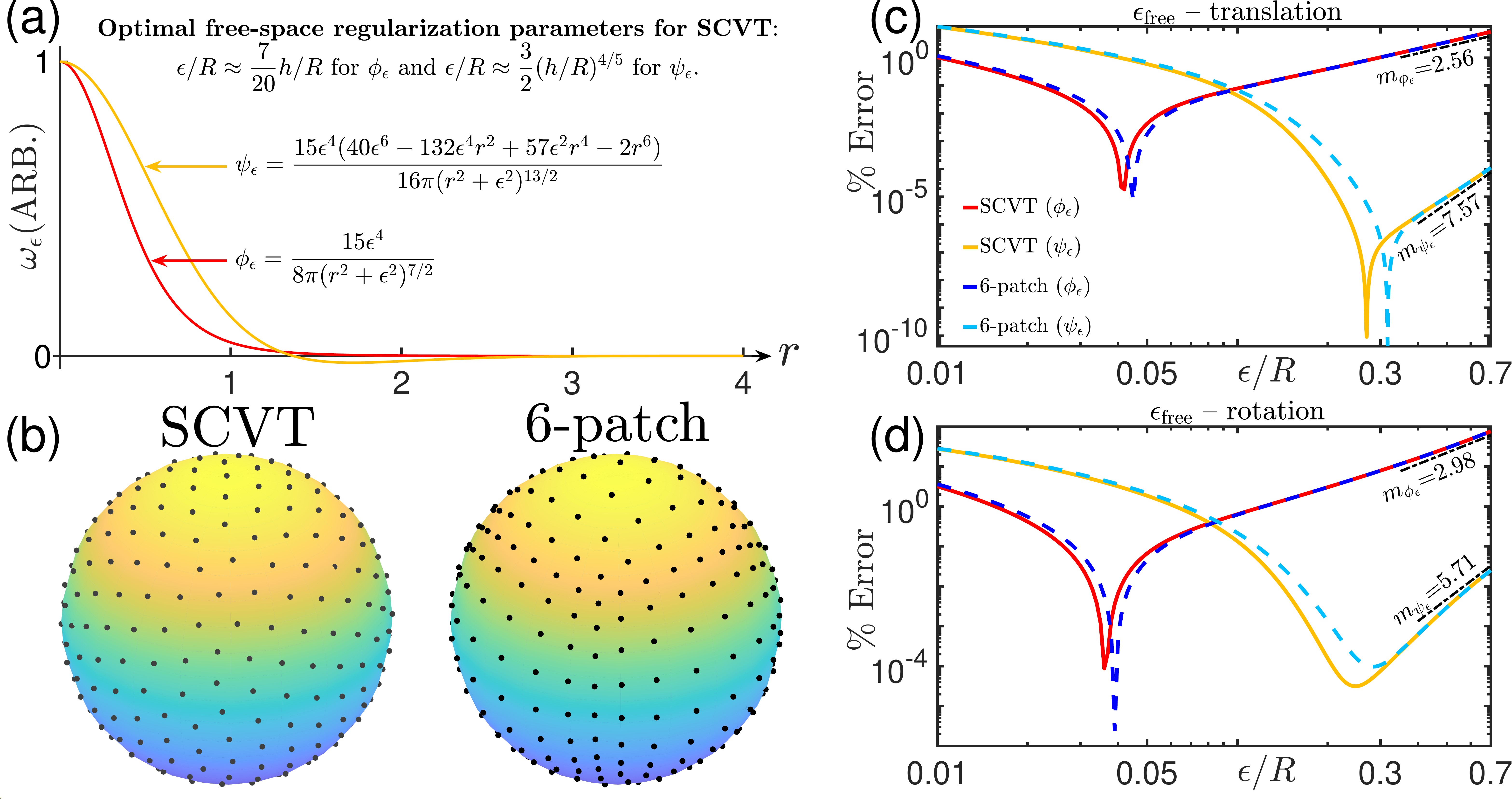}
    \caption{Finding optimal regularization parameters in free space: (a) Regularization functions $\phi_\epsilon$ and $\psi_\epsilon$ are normalized such that their values at the center are equal; (b) Surface discretizations ($N$ = 296) on a sphere using SCVT and 6-patch showing the point distributions. Optimal regularization parameters for (c) translation and (d) rotation in free space with $N=1000$ (SCVT) and $N$ = 1016 (six patch): the minima of the percent errors for the four different combinations are used to identify the value of the optimal regularization parameters.  The decay properties of the regularization functions $\phi_\epsilon$ and $\psi_\epsilon$ are shown as black dash-dot lines that represent the slope of the best fit line used to determine the exponent $m$ in Eq. \eqref{eq:total_err} as $\epsilon \to \infty$.}
    \label{fig:optimal_blob_sizes}
\end{figure}

 In our work, we choose $\omega_\epsilon(r)$ as either:
$$\ds
    \phi_\epsilon(r) = \frac{15 \epsilon^4}{8 \pi (r^2 + \epsilon^2)^{7/2}}\\
\qquad\text{or} \qquad
    \psi_\epsilon(r) = \frac{15 \epsilon^4(40 \epsilon^6 - 132 \epsilon^4 r^2 + 57 \epsilon^2r^4 - 2r^6)}{16 \pi (r^2 + \epsilon^2)^{13/2}},
$$ where $r = |\mathbf{x} - \mathbf{x}_k|$ and $\epsilon$ represents the regularization parameter. Two-dimensional plots of $\phi_\epsilon(r)$ and $\psi_\epsilon(r)$ are shown in Figure \ref{fig:optimal_blob_sizes}(a), for reference.  The function $\phi_\epsilon$ is positive for all $r$, but $\psi_\epsilon$ must have positive and negative values to satisfy the moment conditions in Eq.~\eqref{bealeconditions},  which increases the order of accuracy in GSIRS, see Sec. \ref{sec:res_regfs} and App. \ref{app:num_sims}. Both functions have integral equal to one, but $\psi_\epsilon$ must have a wider positive region to compensate for the negative contribution to the integral, which is also reflected in the size of optimal regularization parameter found in Sec. \ref{sec:res_regfs}. 
 We used the Stokes flow values of drag and torque on a sphere to optimize the value of a regularization parameter, $\epsilon$, in the MRS.

We minimize discretization error by analyzing the six-patch and the SCVT discretization schemes, see Figure \ref{fig:optimal_blob_sizes}(b). The six-patch method distributes points from a bounding cube to the sphere's surface \cite{ainley2008method,cortez2005method,smith2018nearest, smith2009boundary}. The number of points, $N$, is determined using the formula $N = 6n^2 - 12n + 8$, where $n$ is the number of points on each edge of the cube. The six-patch method is commonly used and is easy to implement, but the point distribution is not uniform on the sphere's surface, which may cause errors in the simulation. 

Our SCVT simulations used the method of Du et al. (2003) \cite{Du2003} to produce a nearly uniform point distribution on the sphere using the SCVT software available at Lili Ju's website \cite{LJSCVT} with a constant density function. Figure \ref{fig:optimal_blob_sizes}(b)  compares the uniformity of the point distributions for the SCVT and 6-patch methods with $N = 296$ points. Figure \ref{fig:optimal_blob_sizes}(c)-(d) shows the percent errors relative to theory versus regularization parameter size for a translating sphere and a rotating sphere in free space computed using the MRS for both discretization types.

\section{Results}\label{sec:results} 

\subsection{\label{results:expts} Comparison of experiments with theory}

Figure \ref{fig:drag_torque} shows the close agreement between experiments and the theory of Lee and Leal \cite{Lee_Leal_1980sphere_motion}, which provide a level of precision unmatched by previous experimental tests of theory. The correspondence relied critically on removing any significant effects of other boundaries. The tank was large enough (approximately $48R\times 48R\times 30R$) that the predicted torque increase was negligible and ignored. However, theory predicts a significant increase in the drag on a sedimenting sphere $(\approx10\%)$ from the other five boundaries. Other unknown finite size effects such as corners, free surface motion, etc. were small enough that the theory was still accurate in a tank experiment of the size we used.

\subsection{\label{results:sims} Optimizing the MRS and GSIRS}\label{sec:results_opt}

\subsubsection{\label{sec:res_regfs} Comparison of regularization functions}

We use plots of percent error versus regularization parameter size to assess the order of accuracy of each regularization function finding the exponent $m$ in Eq. \eqref{eq:total_err} from the slope of a log-log plot of percent error versus $\epsilon$, see Figure \ref{fig:optimal_blob_sizes}(c)-(d).  Using the regularization function $\psi_\epsilon$ instead of $\phi_\epsilon$ increases the slope from $m=$ 2.56 to $m=$ 7.57 in the translation data and from $m=2.98$ to $m=5.71$ in the rotation data, where the number of discretization points was $N_\text{six-patch}=1016$ and $N_{\rm SCVT}= 1000$. Thus, \SE\ shows a higher order of accuracy than \PE\, which we attribute to the former satisfying the higher moment conditions, as described in Sec. \ref{sec:sims} and App. \ref{app:num_sims}.

These plots also allowed us to find the optimal regularization parameters for each type of motion and each combination of discretization method and regularization function.  The value of $\epsilon$ that minimizes the percent error was easily identifiable, as shown in Figure \ref{fig:optimal_blob_sizes}(c)-(d). The optimal value of $\epsilon$ for the SCVT simulations varies smoothly as a function of the average discretization size,  $\ds h=\sqrt{\frac{4\pi R^2}{N}}$, so we found fit functions that relate $\epsilon$ and $h$ for each of the regularization functions (see Figure \ref{fig:fitting_optimal_blob_sizes} in App. \ref{app:num_sims} for details): 

\begin{align}
\epsilon_{\rm est}/R&\approx \frac{7}{20}h/R \,\, \text{ for } \phi_\epsilon  \quad \text{ and } \quad  \epsilon_{\rm est}/R\approx \frac{3}{2} (h/R)^{\frac{4}{5} } \quad \text{for } \psi_\epsilon \label{eq:ep_fit}.
\end{align}
These formulae allow other researchers to bypass the free-space-optimization process and use the fit functions instead.
% to find the optimal value of $\epsilon$. 
The six-patch data has a nonuniform discretization length, so no comparable formulae are possible. The value of $\ds \epsilon_{\rm est}/R$ for $\phi_\epsilon$ is very different than the value found for cylinders by Shindell et al. (2021)\cite{shindell2021bacmotilitysurface}. In their work, they found that $\epsilon/R = (h/R)/6.4$, where $h$ was the discretization length between Stokeslets on a side surface of a cylinder. Thus, we confirm the  Shindell et al. result that the relationship between $h$ and $\epsilon$ depends on the geometry of the solid object being simulated.

\subsubsection{Comparison of discretization methods}
\begin{figure}[ht]
    \centering
    \includegraphics[width=1.0\textwidth]{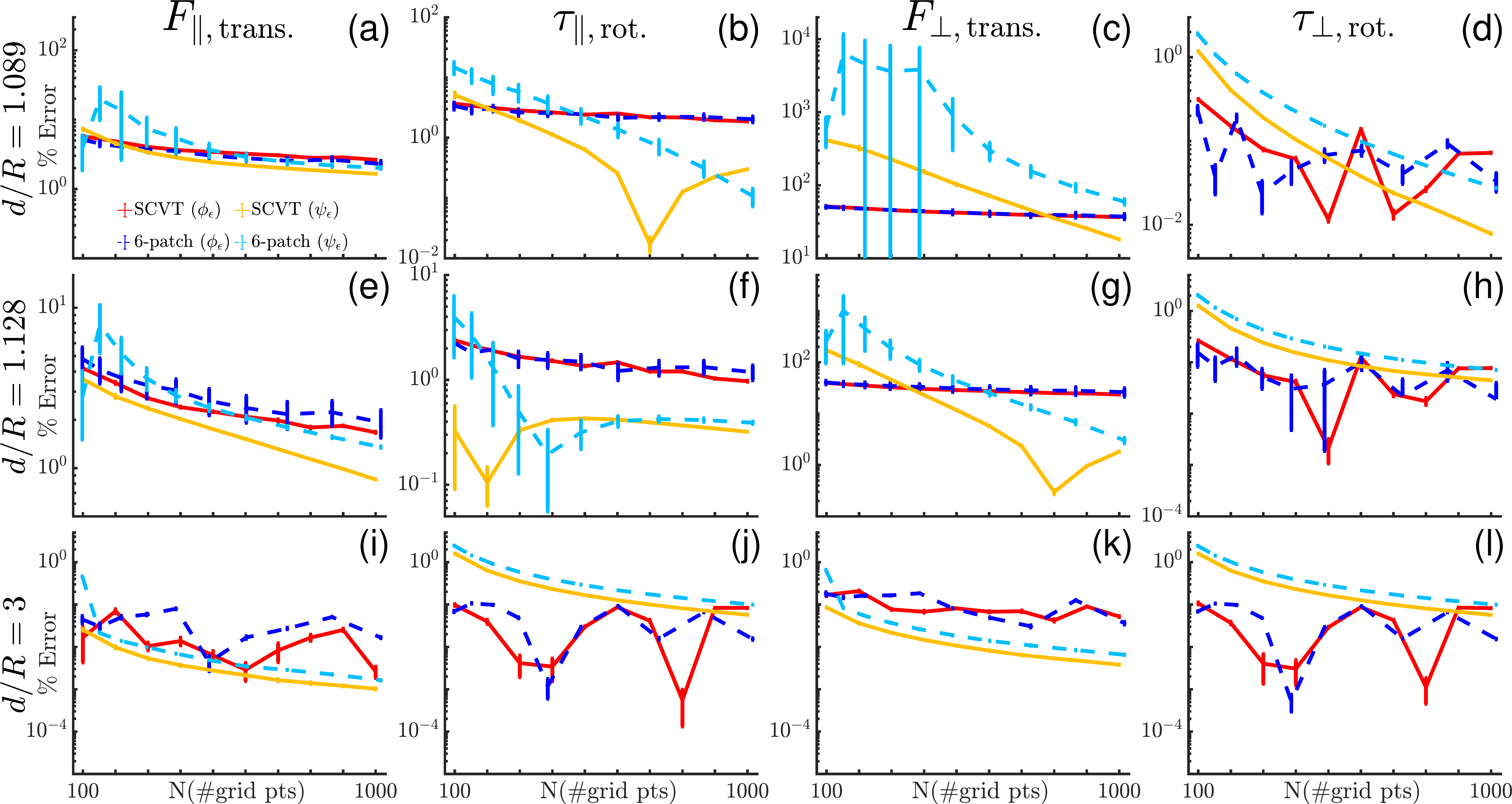}
\caption{Percent errors relative to theory for the four fundamental sphere motions versus number of discretization points, $N$, at three locations near a boundary: $d/R =1.089 $ (top row), $d/R = 1.128$ (middle row), and $d/R = 3$ (bottom row): (a), (e), and (i) are drag when moving parallel to the boundary, (b), (f), and (j) are torque with the rotation axis parallel to the boundary, (c), (g), and (k) are drag when moving perpendicular to the boundary, and (d), (h), and (l) are torque with the rotation axis perpendicular to the boundary.  Solid lines are SCVT simulations and dashed lines show six-patch data. Red and dark blue curves are simulations using the $\phi_\epsilon$ regularization function, whereas the cyan and gold curves used the new $\psi_\epsilon$ regularization function.  Each motion was simulated 36 times with the discretization rotated along a random axis and a random angle. 
Percent error is the mean of the 36 random trials, and error bars are the standard deviation. The SCVT simulations show a lower variance when compared with the six-patch simulations (see Figure \ref{fig:RSD}), and the SCVT generally show smaller percent errors at the highest values of $N$. The top row shows that once the gap between the edge of the sphere and the boundary, $d/R - 1 = 0.089$, is smaller than a discretization length, $h/R \in [0.11, 0.35]$ for $98 \le N \le 1016$, the percent error and the variance are much higher. 
}
    \label{fig:percent_error_near_wall}
\end{figure}

 We used the GSIRS, the two regularization functions, $\phi_\epsilon$ and $\psi_\epsilon$, and the free-space-optimized regularization parameters found above to simulate drags and torques for the four fundamental motions: perpendicular translation, parallel translation, perpendicular axis rotation, and parallel axis rotation. We used a range of discretization points $100\leq N\leq 1016$ and rotated the discretized sphere with respect to the boundary to assess the importance of uniformity in the point distribution.  We evaluate the performance of the optimized six-patch and SCVT simulations near the boundary ($d/R \in 1.089, 1.128, 3.0$) in Figure \ref{fig:percent_error_near_wall} because both types perform well far from the boundary. 

The SCVT simulations (solid curves) generally outperform the six-patch method (dashed curves), especially as the number of grid points increase. The six-patch simulations show relative standard deviations (RSD) as high as 90\% when the discretization is randomly rotated, while the SCVT has an RSD of 1-2\% across all types of motion near the boundary, as shown in Figure \ref{fig:percent_error_near_wall} (See Figure \ref{fig:RSD} and App. \ref{app:num_sims} for more details). The asymmetry of the six-patch method therefore requires care to ensure the discretization does not skew results near the boundary, whereas the SCVT shows very little variance in the computed force or torque. However, we note that the simulations for the closest boundary distance $d/R=$ 1.089 shown in Figure \ref{fig:percent_error_near_wall} (a)-(d) show large variance and large percent errors even at the highest resolution for all simulation types, which is discussed in the next section.
  
\subsubsection{Comparison of the optimized simulations with experiment and theory} 

\begin{figure}
\begin{tabular}{cc}
\includegraphics[width=0.45\columnwidth]{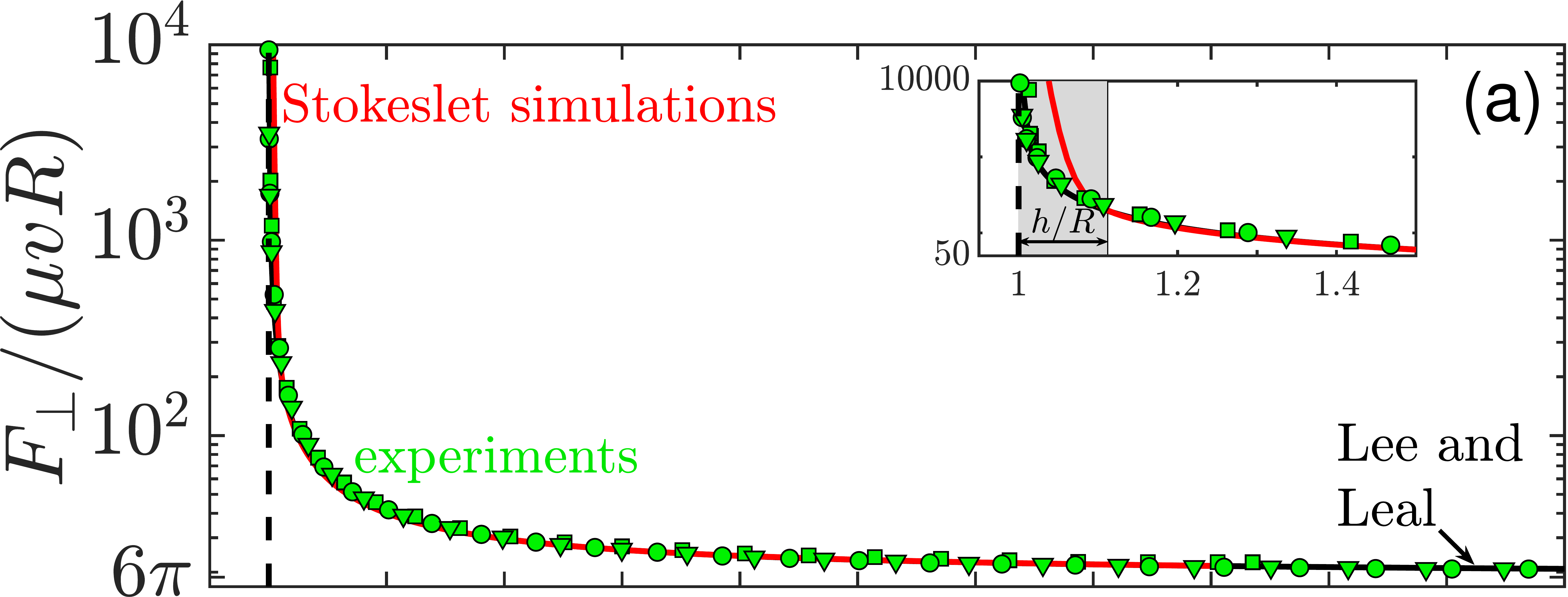}\hspace{0.25cm}
    & \includegraphics[width=0.45\columnwidth]{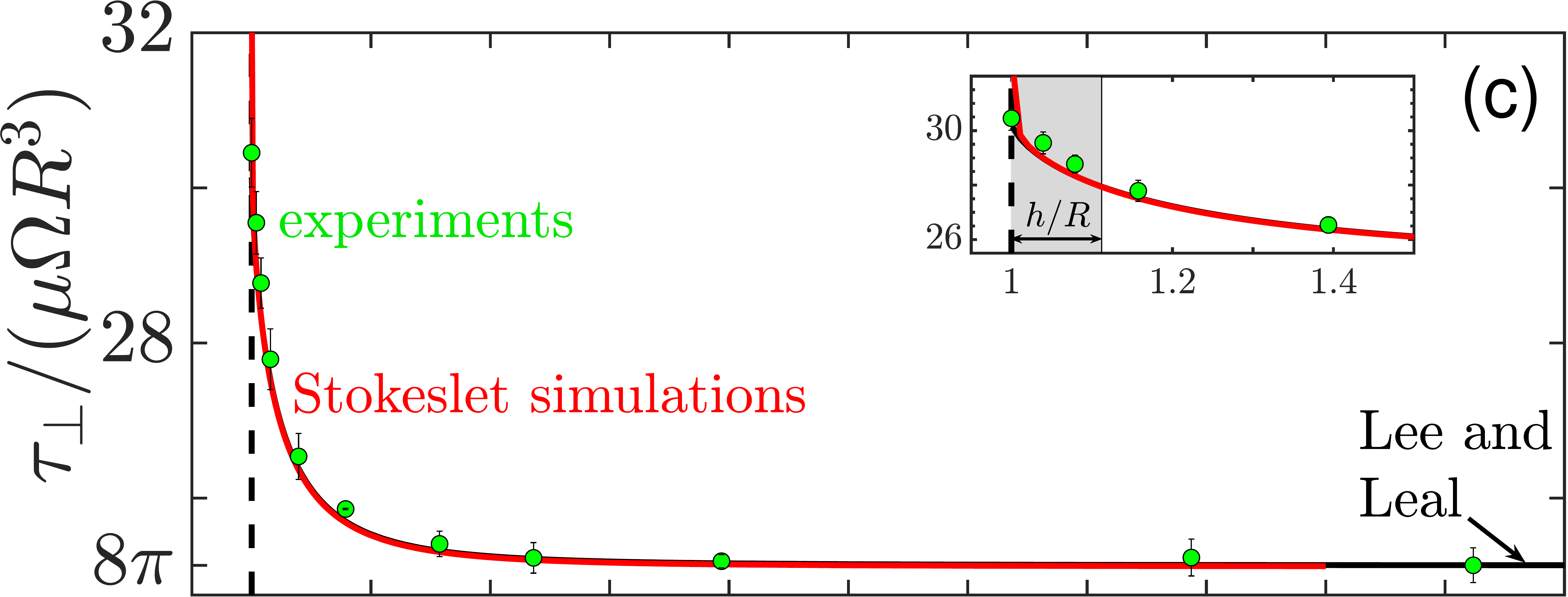}\\
    \hspace{0.075cm}\includegraphics[width=0.45\columnwidth]{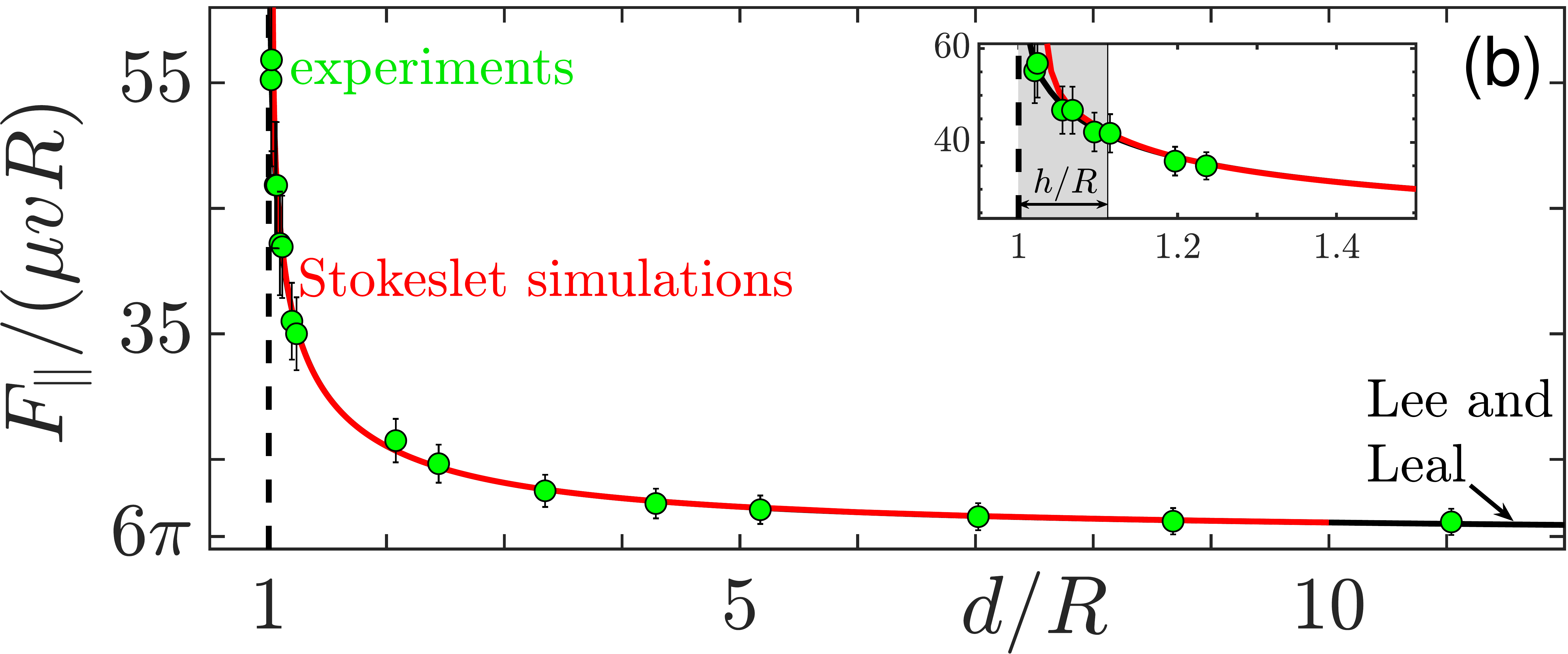} \hspace{0.25cm}
    & \includegraphics[width=0.45\columnwidth]{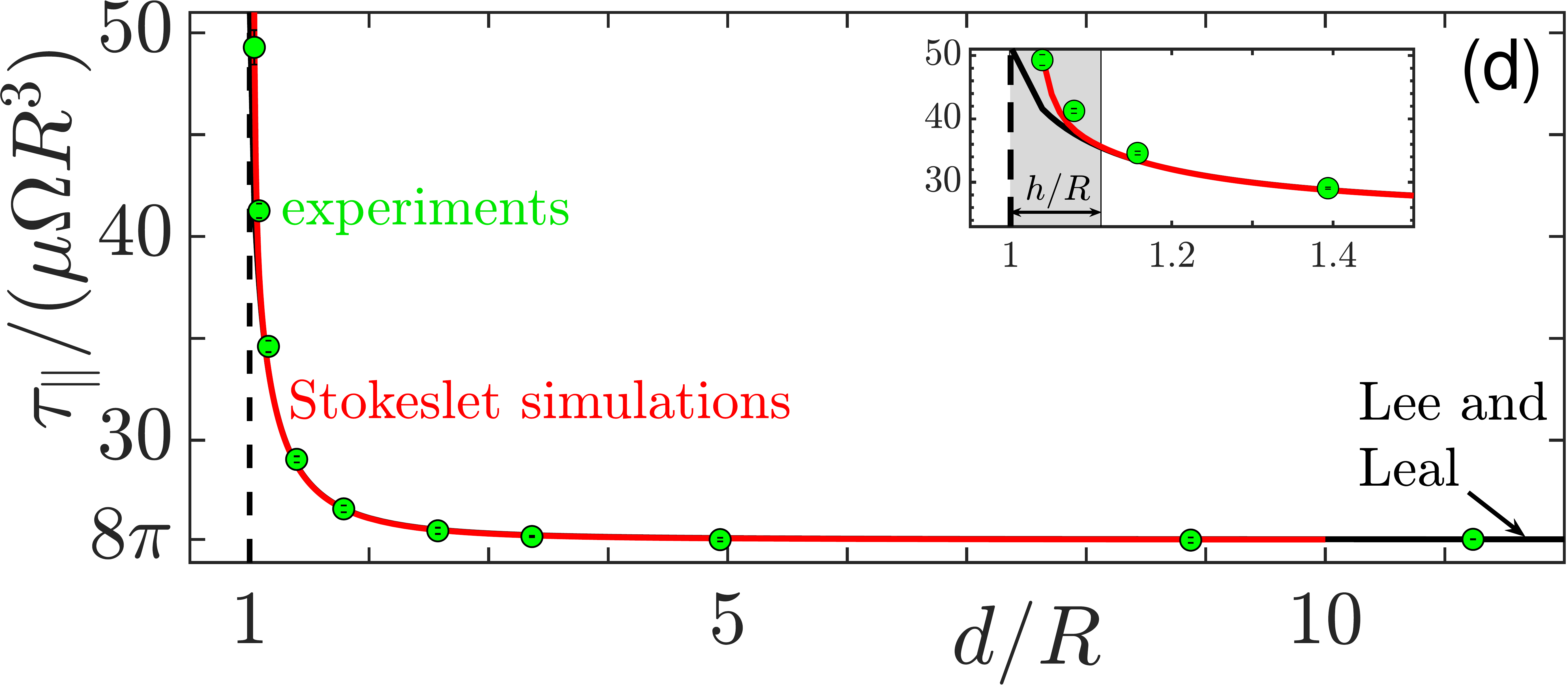} \\
\end{tabular}
\caption{Dimensionless drag and torque versus scaled boundary distance, $d/R$,
on a sphere, where $R$ is the radius of the sphere, and $d$ is the distance from the center of the sphere to the boundary. The solid black curves shows the theory by Lee and Leal \cite{Lee_Leal_1980sphere_motion}, the solid red curves are optimized GSIRS simulations, and the solid green symbols are dynamically similar experiments. Panel (a) shows translation perpendicular to a boundary, $F_\perp/(\mu v R)$ and (b) is motion parallel to a boundary $F_\parallel/(\mu v R)$, scaled using $\mu$, the dynamic viscosity, $v$ is the speed, and $R$. Panels (c) and (d) show dimensionless torque versus boundary distance for rotation with the axis perpendicular to the surface $\tau_\perp/(\mu\Omega R^3)$ and with the rotation axis parallel to the boundary $\tau_\parallel/(\mu\Omega R^3)$. Experiments, simulations, and theory all agree very well outside of the near boundary region. The experiments agree with theory at most distances except for $\tau_\parallel$, which diverges from theory near the boundary. We attribute this difference to  locating the edge of the sphere (see text). However, the insets show the simulations diverge from theory in the near boundary region for three of the four  motions, which occurs when the edge of the sphere is closer than the average discretization size ($h/R = 0.11$) to the boundary. The inset in panel (c) shows that the simulations perform better for $\tau_\perp$ than the other three motions.}\label{fig:drag_torque}
\end{figure}

We used the higher-order regularization function, $\psi_\epsilon$, the SCVT discretization method ($N = 1000$), and the free-space-optimized value for the regularization parameter to simulate sphere motion for a large range of boundary distances for comparison with the experiments and theory, as shown in Figure \ref{fig:drag_torque}.

The perpendicular translation, $F_\perp$, data in Figure \ref{fig:drag_torque}(a) shows that the data all agree for most boundary distances, except very near the boundary, which is consistent with what we found in Figure \ref{fig:percent_error_near_wall}. 
Near the boundary, the dimensionless drag increases by a factor nearly $10^4$, and the distance moved between frames in the experiment is around $10 \mu$m, which highlights the precision needed in the motion tracking.  However, the inset in Figure \ref{fig:drag_torque}(a) shows that the numerical simulations diverge systematically from the theory and experiments when the sphere is near $d/R\approx1.1$. We find the same systematic divergence from theory when using either \PE\ or \SE\, and find a similar result for other motions, as described below. 

Figure \ref{fig:drag_torque}(b) shows that the parallel translation, $F_\parallel$, simulations perform better overall than the simulations of perpendicular translation, but similarly, the inset shows deviation from theory in the simulations at the about $d/R=1.1$. The maximum drag increase expected here is a factor of about  ten compared to about $10^4$ in the perpendicular translation case, which may explain why the deviation is less pronounced until about $d/R=1.05$.

Figure \ref{fig:drag_torque}(c) shows that the three data types: theory, experiments, and simulations all agree very well for torque with the rotation axis perpendicular to the boundary, $\tau_\perp$. The deviation of the simulations from experiments and theory occurs much closer to the boundary and is smaller. However, $\tau_\perp$ shows the least increase near the boundary among the four motions studied.

Torque with the rotation axis parallel to the boundary, $\tau_\parallel$, is shown in Figure \ref{fig:drag_torque}(d). The experiments show some systematic deviation from theory, which we attribute to uncertainty in establishing when $d/R=0$ because the precession of the sphere (0.25mm) is  the same order  as the closest boundary distance ($d=0.5$mm). The simulations also show systematic deviation from theory at about $d/R=1.1$ in these data.

In summary, none of the simulation types perform well when the sphere is located very near the boundary, see insets in Figure \ref{fig:drag_torque}. Other researchers such as Zheng et al. (2023) \cite{ZHENG2023112} have suggested that the method of images does not perform well when the gap between an object and the wall is smaller than the size of the regularization parameter $\epsilon$. In our data, the normalized gap where numerical simulations diverge from theory ($h/R = 0.11$) is larger than the optimal value of $\epsilon/R=0.042$ for \PE\ and smaller than the optimal value of $\epsilon/R = 0.27$ for \SE. Instead, we find that the GSIRS simulations 
diverge from theory when the gap size is within the average discretization size. Appendix \ref{app:near_opt} discusses options for minimizing the error very near the boundary, but our result gives researchers an empirical rule to identify boundary distances where the GSIRS may not be accurate, i.e. when $d/R<1+ h/R$.

\section{Discussion}\label{sec:discussion} 

The goal of this work is to develop methods to precisely calibrate the method of images for regularized Stokeslet simulations of spheres moving near a boundary. We used measurements of forces and torques from macroscopic experiments to verify the theory by Lee and Leal \cite{Lee_Leal_1980sphere_motion} before using it for optimization. We chose this theory because it includes all four motions though other analyses give the same predictions, see for example \cite{Brenner_1961, oneill1964spherenearwall, Dean_ONeill_1963_rotating_sphere}. Our results provide the first comprehensive experimental test of these theories, and Figure \ref{fig:drag_torque} also shows that the theory can be used to effectively remove the finite size effects in an experiment. Considering the significant finite size effects ($\approx 10\%$) and that the tank has corners, etc., the theory worked very well despite its underlying assumptions of zero Reynolds number and an infinite plane boundary.  

Using the verified theory, we developed computational methods to accurately model a sphere moving near a boundary. Our method of finding the optimal regularization parameter, $\epsilon_{\rm free}$, provides a computationally efficient way to optimize the GSIRS, whether using the \PE\ or the \SE\ regularization function, though we find the latter to be a better choice because of its stability, convergence, and error properties, as shown in Figure \ref{fig:percent_error_near_wall}. The fit functions in Eq. \eqref{eq:ep_fit} allow researchers to bypass the optimization process and set a value of $\epsilon$ when using either \PE\ or \SE . 

Our results also confirm the finding of Shindell et al. (2023) \cite{shindell2021bacmotilitysurface}; the relationship between discretization size and regularization parameter depends on the shape of the simulated object. Thus, using a value for the regularization parameter that is not optimized for a given geometry introduces unknown errors into MRS and GSIRS simulations.

We also show that the SCVT discretization method is superior to the six-patch method. The asymmetry of the six-patch point distribution causes significantly increased variance in the computed values when the sphere is rotated with respect to the boundary (see Figure \ref{fig:percent_error_near_wall} and Figure \ref{fig:RSD} in the Appendix).
Thus, a discretization method that creates a more uniform point distribution should be chosen when simulating a sphere moving near a boundary. 

We found that the GSIRS lost accuracy when the sphere was near the boundary (see the insets in Figure \ref{fig:drag_torque}){. Zheng et al. (2023) \cite{ZHENG2023112} commented that the image system may suffer inaccuracy within a distance of order of the regularization parameter from nonzero values of the image sources crossing the boundary. Our GSIRS simulations using \PE\,the same regularization function as used by Zheng et al., show high percent errors from theory when the distance of the edge of the sphere (0.089R) was larger than the size of the regularization parameter ($\epsilon = 0.042R$) (see the top row of Figure \ref{fig:percent_error_near_wall}). Furthermore, we found large differences relative to theory at the same distance when using \SE, and in this case, the gap is much smaller than the size of the regularization parameter, $\epsilon = 0.27R$. The force distribution is modified near the wall due to the presence of the image system, but we 
find that a good empirical rule for establishing the number of grid points needed is to determine the closest distance to the boundary that will be simulated and ensure that the average discretization length is smaller than this distance. Our method of optimizing the simulations in free space and this discretization rule will give better results than using the size of the regularization parameter to establish the limits of accuracy. 

In conclusion, using the GSIRS with a uniform point distribution such as the SCVT, a high-order regularization parameter function such as $\psi_\epsilon$, and a free-space-optimized regularization parameter provides an efficient and accurate method for simulating spheres moving near a boundary providing the edge of the sphere is kept outside of the average discretization length. These results along with our experimental validation of theory and MATLAB and PYTHON implementations of the Lee and Leal theory provide important resources for other researchers using the GSIRS or other simulation methods to assess the accuracy of their numerical results. Future study should explore how our optimization strategy and calibration process work for other geometries including compound objects such as a sphere with a helical flagellum.

\section*{Acknowledgments}
This research was supported by the collaborative NSF grant through the Physics of Living Systems (PHY-2210609 to H.N., A.G., O.S., and F.H., and PHY-2210610 to B.R., K.B., and J.M.) and the collaborative NSF grant through the DMS/NIH-NIGMS Initiative to Support Research at the Interface of the Biological and Mathematical Sciences (DMS-2054333 to R.C. and DMS-2054259 to H.N. and A.G.). We thank Trinity University for the provision of computational resources and NSF MRI-ACI-1531594 for providing Trinity the High Performance Scientific Computing Cluster. We thank the Centre College Faculty Development Fund and Gary Crase for assistance with experimental equipment and design.
% \end{acknowledgments}

\section*{Author Information}

\subsection{Authors and Affiliations}

\noindent\textbf{Physics Program, Centre College of Kentucky, Danville, KY 40422, USA}

Kathleen Brown, Jonathan McCoy, and Bruce Rodenborn

\noindent \textbf{Department of Mathematics, Tulane University, New Orleans, LA 70118, USA}

Ricardo Cortez

\noindent \textbf{Department of Mathematics, Trinity University, San Antonio, TX, 78212, USA}

Hoa Nguyen and Amelia Gibbs

\noindent \textbf{Department of Physics and Astronomy, Trinity University, San Antonio, TX, 78212, USA}

Orrin Shindell

\noindent \textbf{Department of Biology, Trinity University, San Antonio, TX, 78212, USA}

Frank Healy

\subsection{Contributions}
B.R., H.N., R.C., O.S., and F.H. were involved in conceptualization of the project. For the original draft, B.R. wrote, reviewed, and edited the manuscript while H.N. and R.C. wrote the Numerical Methods section. B.R., H.N., R.C., and O.S. worked together on revising the final draft. B.R. designed the experiments and analyzed the results. B.R. and O.S. created the numerical implementation of theory in MATLAB. B.R. mentored two undergraduate students (K.B. and J.M) to conduct experiments, analyze the results, and assist in developing the supplementary information including a PYTHON version of the theory. H.N. designed the numerical codes and mentored one undergraduate student (A.G.) on model development and implementation to generate simulation results and data analysis. A.G. conducted simulations and analyzed the results. R.C. provided the regularization function $\psi_\epsilon$ and insights into the GSIRS. All authors reviewed the manuscript.

\section*{Additional information}

\textbf{Competing interests} 

The authors declare no competing interests.

\section*{Data Availability}

The data presented in this manuscript are provided via GOOGLE DRIVE at \url{https://drive.google.com/drive/folders/1w-U42eVbh17HiOrBwUJB6SW0ttMy6n0s?usp=share_link}. Other data and numerical simulation code that support the findings of this study are available from the corresponding author upon reasonable request.

%\bibliography{mainNotes.bib}% Produces the bibliography via BibTeX.

%

\appendix
\section{Numerical simulations}\label{app:num_sims}

\subsection{Simulation details}

We used three sets of simulations to minimize the error relative to theory when simulating the sphere motions we studied in our experiments. The first set of simulations used the MRS to understand which regularization function, $\phi_\epsilon$ or $\psi_\epsilon$, converges most rapidly as the regularization parameter size increases. These simulations were also used to find an optimal value for the regularization parameter $\epsilon$ for each combination of the regularization function and discretization method. The second set of simulations used GSIRS and the optimal regularization parameters from the first set of simulations to assess how orientation relative to the boundary of the six-patch and SCVT discretizations affects the percent error when compared to theory. We then simulated sphere motion for a large range of boundary distances using our free-space-optimized regularization parameters, SCVT, and the high-order regularization function $\psi_\epsilon$ for comparison with the experiments and theory. For the last set of simulations, we further explored how to optimize values for the regularization parameter at very small boundary distances for the sphere motions.
\begin{table}[H] 
\caption{Parameters values used in numerical simulations.} 
\label{table_num_params}
\begin{center}
\begin{tabular}{ll}
\toprule
\textbf{Parameter}& \textbf{Value}  \\
\hline
\midrule
Dynamic viscosity & $\mu = 10^{-3}\, \rm{Pa\cdot s}$ \\
\midrule
Sphere radius & $R =1\mu \rm{m}$ \\
\midrule
Translation speed &$U=1\mu \rm{m}/ s$  \\
\midrule
Rotation rate & $\Omega/(2\pi) = 154\, \rm{Hz}$  \\
\midrule
{Discretization \#} & $N$(six-patch) $ \in \{98, 152, 218, 296,  $\\
&$386, 488, 602, 728, 866, 1016\}$\\
\midrule
& $N$(SCVT) $\in$ \{100:100:2400\} \\
\bottomrule
\hline
\end{tabular}
\end{center}
\end{table}

We incremented the regularization parameters from $\epsilon/R = 0.01$ to 0.7 in steps of 0.001 for each motion, each type of regularization function, and each discretization method.  
We varied the regularization parameter for each type of motion and found the value of $\epsilon$ that minimized the difference between the simulation and theory.  
The range of simulation parameters used are reported in Table \ref{table_num_params}.
We then computed the percent errors between the simulations and theory versus regularization parameter size for each motion, see for example Figure \ref{fig:optimal_blob_sizes}(c)-(d). Overall, 55,280 calculations were conducted: two motion types (translation and rotation), two regularization functions ($\phi_\epsilon$, $\psi_\epsilon$), two discretization methods (SCVT, 6-patch), 10 values of $N$, and 691 values of regularization parameters, $\epsilon$.

\subsection{Numerical errors}

Given a point force at $\mathbf{x}_0$, the MRS derived for an arbitrary regularization function, $\omega_\epsilon(r)$ where $r = |\mathbf{x} - \mathbf{x}_0|$ is written in terms of the functions $\Gd(r)$ and $\Bd(r)$, which satisfy  $\Delta\Gd = \omega_\epsilon$ and $\Delta\Bd = \Gd$. In the far field, these functions approach their singular  counterparts $G_0(r) = -(4\pi r)^{-1}$ and $B_0(r) = -(8\pi)^{-1} r$. The far field error can be controlled by constructing a regularization function that produces rapid convergence of the regularized Stokeslet to the singular one. The convolution of the regularization function $\omega_\epsilon(\bx)$ with a smooth function $f(\bx)$ can be a high-order approximation of $f(\bx)$ by forcing the regularization function to satisfy certain moment conditions.  Nguyen et al. \cite{nguyen2014reduction} show that the convergence rates of these approximations are directly related to the decay rate of $\omega_\epsilon(r)$ when the following moment conditions are satisfied:
\begin{eqnarray}\label{blobmoment0}
\int_0^\infty r^2 \omega_\epsilon(r) dr &=& (4\pi)^{-1} \\ 
\int_0^\infty r^4 \omega_\epsilon(r) dr &=& 0.\label{blobmoment1}
\end{eqnarray}

Nguyen et al. \cite{nguyen2014reduction} also established that if
$\omega_\epsilon(r)$ does not satisfy condition~\eqref{blobmoment1}, then $|\Bd(r)-B_0(r)| = O(r^{-2})$ for $\epsilon\ll r$ regardless of the decay properties of $\omega_\epsilon(r)$. 

The sphere motion we simulate generates forces distributed over surfaces of solid objects rather than volumes, so
the  conditions that reduce the regularization error are  different. The velocity field at any location is given by the surface integral of the Stokeslet, 
$\ds \int_\Gamma S(\bx,\by(A)) \ff(A) dA$. Approximating this integral by a quadrature and replacing the Stokeslet with a regularized Stokeslet, $\ds \sum_{k=1}^N S_\epsilon(\bx,\by_k) \ff_k \Delta A_k$ creates two types of error: one from the discretization of the integral and the other from the regularization of the Stokeslet. The form of the total error is important to establish convergence of the method. 
The total error is given by:
 \begin{equation}\label{eq:total_err}
Err(\bx) = O\left( \frac{h^p}{\epsilon^q} \right)+ O(\epsilon^m),
\end{equation}
where $h$ is the discretization length and the exponents $p$, $q$, and $m$ depend on the specific implementation of the numerical method used to compute the surface integrals
% ~\cite{cortez2005method},\cite{smith2018nearest},\cite{smith2021richardson} 
\cite{cortez2005method,smith2018nearest, smith2021richardson}. The first term is the discretization error, and its form depends on the regularization function used, the quadrature method, and whether periodicity can be exploited. Our work focuses on the second term, the regularization error, which depends only on the regularization function. Equation \eqref{eq:total_err} implies that the regularization error dominates when $\epsilon$ is large compared to $h$, which we exploit when analyzing the two regularization functions, see Figure \ref{fig:optimal_blob_sizes}(c-d). % Sec. \ref{sec:results}. 

Beale~\cite{beale2001convergent} showed that for surface forces in water waves, the following additional constraint on $\epsilon$ reduces the regularization error by increasing the exponent $m$ in \eqref{eq:total_err}:

 \begin{equation}\label{bealeconditions}
\int_0^\infty ( -4\pi r \Gd(r) - 1) r^{2k} dr = 0, \ \ \ \textrm{ for } k=0,1,2, \dots
\end{equation}

The regularization function $\omega_\epsilon = \fd(r)$ satisfies only criterion \eqref{blobmoment0}, whereas our regularization function  $\omega_\epsilon = \psi_\epsilon(r)$  satisfies~\eqref{blobmoment0}-\eqref{blobmoment1} and~\eqref{bealeconditions} for $k = 0$ and $1$. We show in Sec. \ref{sec:results} that the exponent $m$ for \SE\ is larger than for \PE\, which confirms it's rate of convergence is faster, see Figure \ref{fig:optimal_blob_sizes}, and therefore, \SE\ is a higher order regularization function.

 \begin{figure}
    \centering
    \includegraphics[width=0.65\textwidth]{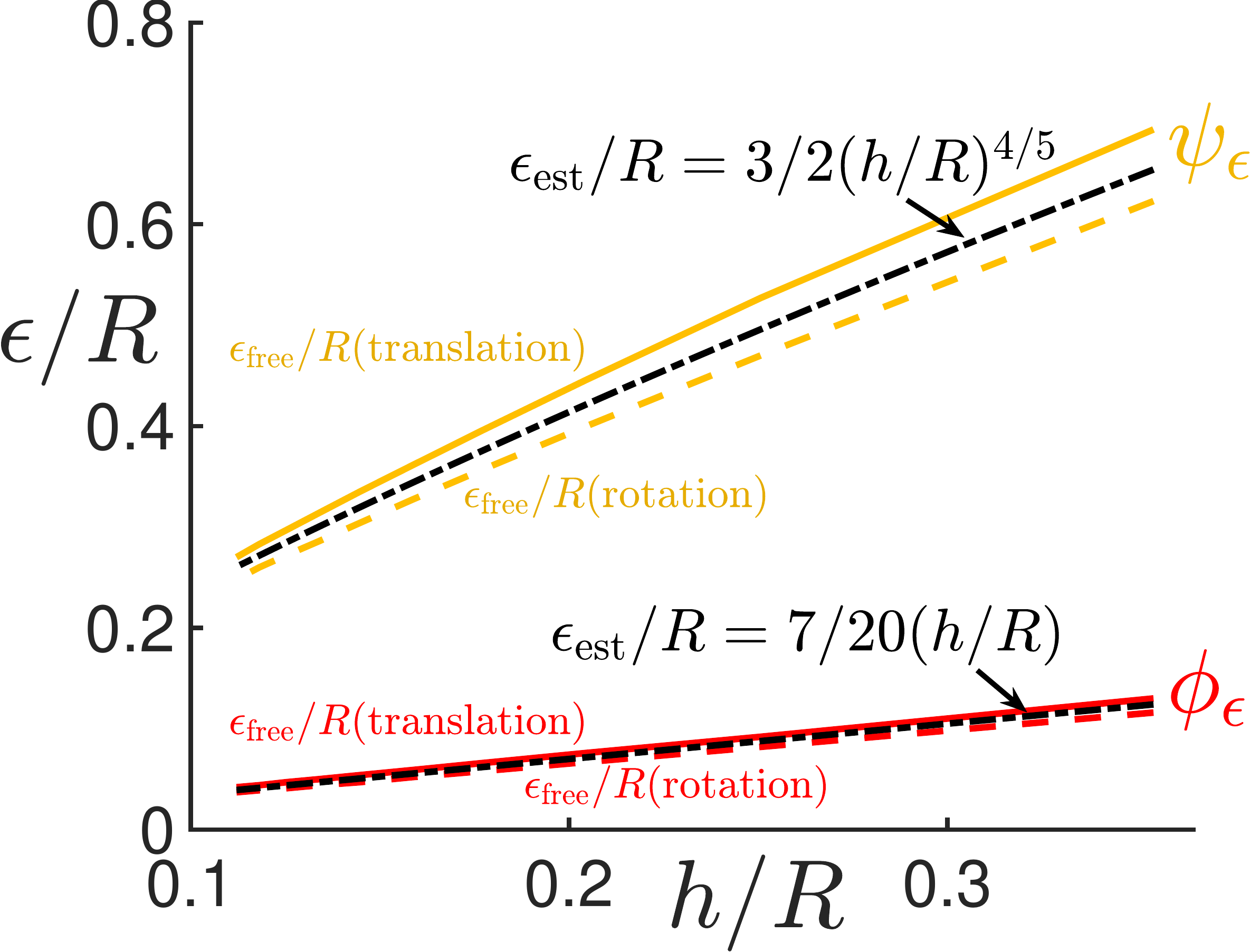}
    \caption{ Optimal regularization parameters in free space for SCVT for different surface discretizations with average discretization lengths, $\ds h=\sqrt{\frac{4\pi R^2}{N}}$, where $N$ is the number of grid points and $R$ is the sphere radius. The optimal regularization parameter, $\epsilon_{\rm free}/R$, for each regularization function can be estimated as a function of the average SCVT discretization length, $\ds h$, for both types of sphere motion. The fit function estimates for the regularization parameters are shown as black dash-dot curves. The relative difference between optimal regularization parameter for rotation and translation is approximately the same for $\psi_\epsilon$ and $\phi_\epsilon$ but is exaggerated for $\psi_\epsilon$ because of the scale in the plot. The mean of the ratio (rotation to translation) of the optimal regularization parameters is 0.91$\pm$0.01 for $\psi_\epsilon$ and 0.88$\pm$ 0.01 for $\phi_\epsilon$.}
    \label{fig:fitting_optimal_blob_sizes}
\end{figure}

\subsection{Discretization errors}

Figure \ref{fig:percent_error_near_wall} shows the mean percent error versus number of grid points at three small boundary distances: $d/R$ = 1.089, 1.128, 3.0. There were 17,280 calculations in total: four motions, two regularization functions, two discretization methods, 10 values of $N$, three distances, and 36 random trials. 

Generally, the percent error decreases as $N$ increases, as expected, but the rate at which this occurs is different. Simulations using $\psi_\epsilon$ (solid gold and dashed cyan curves) show a more rapid decrease in percent error as $N$ increases when compared to matched simulations using $\phi_\epsilon$ (solid red and dashed dark blue curves). The data also show that SCVT simulations using either regularization function (solid gold and solid red curves) generally return a lower mean percent error than the matched six-patch simulations (dashed cyan and dashed dark blue curves) especially when $N\approx$ 1000.  Table \ref{table_percent_errors} lists the range of the percent errors for these simulations when $800 \le N \le 1016$: the performance is especially poor for the translational motion perpendicular to the wall as seen in Figure \ref{fig:percent_error_near_wall} (c). 

Further from the boundary when $d/R = 1.128$, as shown in Figure \ref{fig:percent_error_near_wall} (e) - (h), the data are easier to interpret. The percent errors are reduced typically by an order of magnitude and the variance is much lower, but the SCVT simulations using \SE\ generally converge more quickly, have lower variance, and achieve lower percent error at the highest value of $N$, as shown in panels (e) - (g), though there is no clearly superior method for simulating the perpendicular torque shown in panel (h). Once the distance is $d/R = 3$ the six-patch and SCVT show similar performance in mean percent error values, which are small for all simulation types. 

We extended our analysis of the variance in the different simulation types by plotting the relative standard deviation (RSD) of the mean value of force and torque in Figure \ref{fig:RSD}. The plots of RSD versus number of grid points $N$ show that the SCVT simulations are much less sensitive to the orientation of the sphere with respect to the boundary at any boundary distance.

In Figure \ref{fig:RSD}(c), the six-patch simulations show an RSD as high as 90\% for the perpendicular translation data while the SCVT has an RSD of 1-2\%. The asymmetry of the six-patch discretization method therefore requires care to ensure the discretization does not skew results near the boundary and requires averaging of different orientations for accuracy, whereas the SCVT shows very little variance in the computed force or torque. Thus, averaging over 36 orientations when using an SCVT discretization, as we have done for analysis, is not necessary.

\begin{table}[H] 
\caption{Range of percent errors for $d/R = 1.089$ and $800 \le N \le 1016$ for four motions. The range was determined by finding the smallest (min) and largest (max) percent errors of each combination of the discretization method (SCVT or 6-patch) and the regularization function ($\phi_\epsilon$ or $\psi_\epsilon$) for all these values of $N$.} 
\label{table_percent_errors}
\begin{center}
\begin{tabular}{|c|c|c|c|c|}
        \hline
        min - max & SCVT ($\phi_\epsilon$) &     SCVT ($\psi_\epsilon$) &     6-patch ($\phi_\epsilon$) &     6-patch ($\psi_\epsilon$) \\
        \hline
        $F_{\parallel, \text{trans.}}$ & 2.604 - 2.911 &            1.643 - 1.888 &            1.635 - 3.940 &               1.837 - 3.256\\
        \hline
        $\tau_{\parallel, \text{rot.}}$ & 1.837 - 2.213 &            0.113 - 0.307 &          1.076 - 2.818 &               0.004 - 1.453 \\
        \hline
        $F_{\perp, \text{trans.}}$ & 36.167 - 38.847 &          17.761 - 36.160 &          24.117 - 46.052 &              35.575 - 231.592\\  
        \hline
        $\tau_{\perp, \text{rot.}}$ & 0.023 - 0.076 &          0.007 - 0.018 &       0.005 - 0.137 &            0.0269 - 0.064 \\
\hline
\end{tabular}
\end{center}
\end{table}

  \begin{figure}
    \centering
    \includegraphics[width=1.0\textwidth]{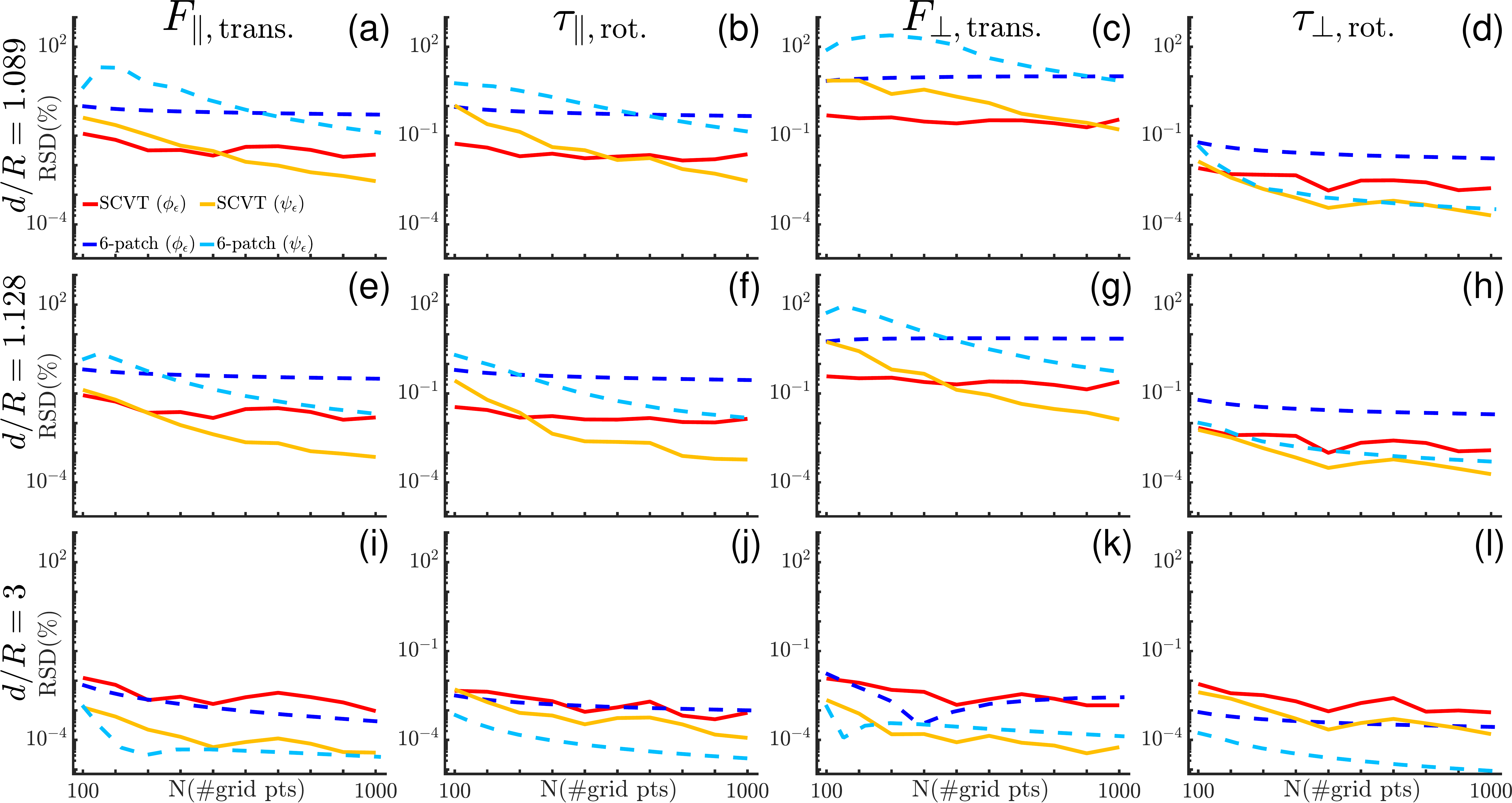}
\caption{The figures show the relative standard deviation (RSD) as a percentage versus number of discretization points, $N$, at $d/R = 1.128$ (top row) and at $d/R = 3$ (bottom row). Panels (a), (e), and (i) are drag for motion  parallel to the boundary, (b), (f), and (j) are torque for rotation with axis parallel to the boundary, (c), (g), and (k) are drag for motion perpendicular to the boundary, and (d), (h), and (l) are torque for rotation with axis perpendicular to the boundary.  Red and dark blue curves are simulations that used the $\phi_\epsilon$ regularization function, whereas the gold and cyan curves used the $\psi_\epsilon$ regularization function.  Each sphere motion was simulated 36 times with the sphere's discretization rotated randomly.  Away from the boundary, all simulation types show a low percent error relative to the theory panels (i)-(l). The SCVT simulations show a lower RSD when compared with the six-patch simulations for most motion types. The SCVT simulations using the $\psi_\epsilon$ regularization function generally show the lowest RSD as $N$ becomes large.}
    \label{fig:RSD}
\end{figure}

\subsection{Minimizing error very near the boundary}\label{app:near_opt}

  \begin{figure}[ht]
    \centering
    \includegraphics[width=.95\textwidth]{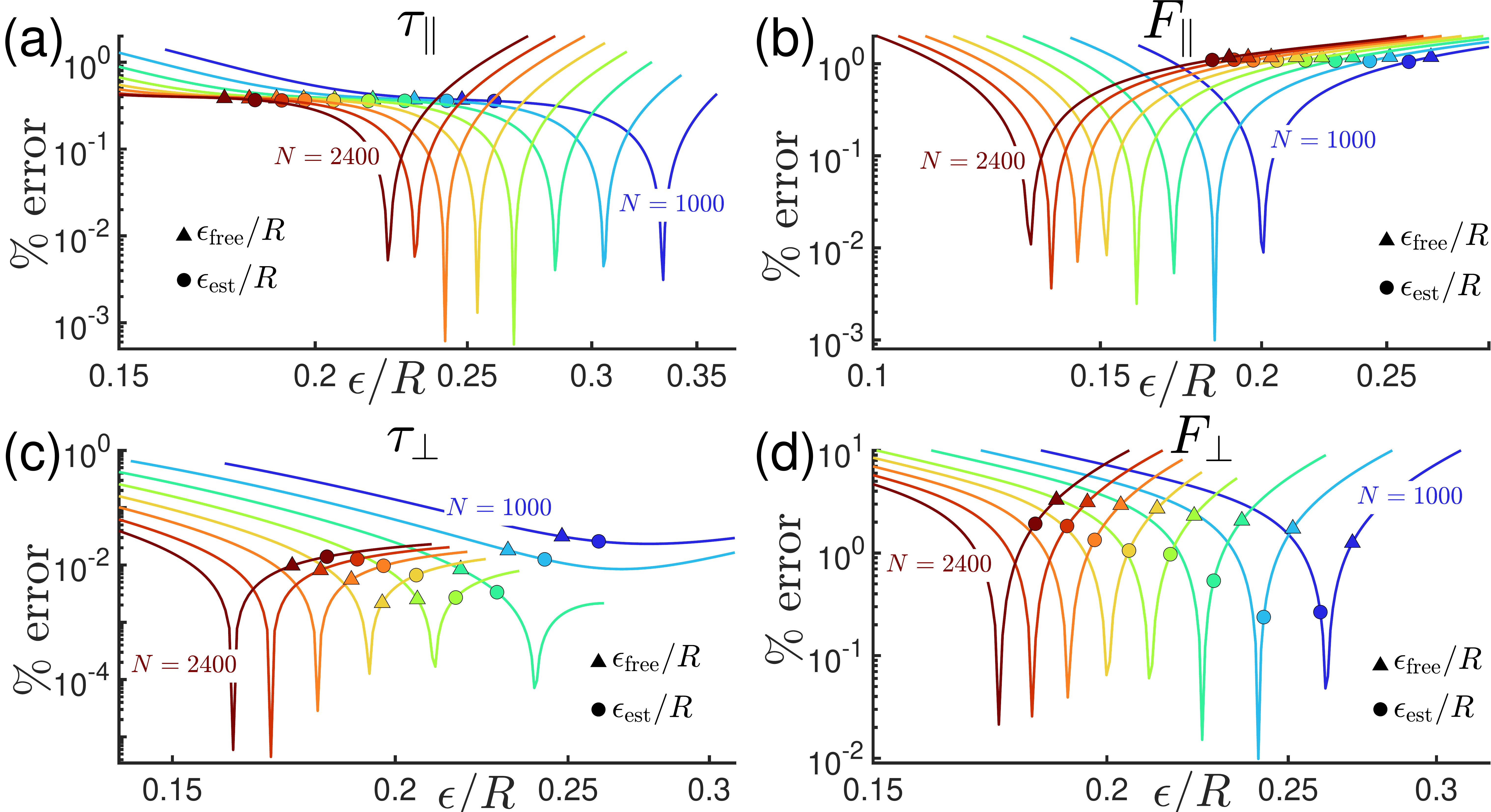}
\caption{Optimizing the regularization parameter $\epsilon$ when the edge of the sphere is one discretization length $h$ from the boundary. The simulations use  the SCVT discretization method and the \SE\ regularization function for eight values of $N$ from 1000 to 2400 in increments of 200 corresponding to eight different discretization lengths $h$. The solid curves are the percent error as the regularization parameter is varied for each value of $N$, which all show a clear minimum that could be used to minimize the difference between simulations and theory, see Figure \ref{fig:epsilon_vs_h_fits}. By contrast, solid triangles are the free-space-optimized values of $\epsilon_{\rm free}/R$ found using the MRS simulations, as described in \textbf{Optimal regularization parameter} section, above. The solid circles are the estimated values of $\ds \epsilon_{\rm est}/R = 3/2(h/R)^{4/5}$ using the fit function as described in Figure \ref{fig:fitting_optimal_blob_sizes}. Using either the free-space-optimized or the estimated value of the regularization parameter introduces a small percent difference relative to theory near the boundary, but slightly better performance further from the boundary.}
    \label{fig:err_vs_eps_all}
\end{figure}

Our approach to optimizing the simulations far from the boundary by using MRS calculations, as described in Sec. \ref{sec:sims}, results in small percent errors for all boundary distances except when the sphere's edge is closer to the boundary than a discretization length (insets of Figure\ref{fig:drag_torque}). 
We investigated this limitation of our optimization method by varying $N \in \{1000:200:2400\}$, and placing the sphere at $d = h + R$ to the wall. We simulated all four motions near the boundary using the $\psi_\epsilon$ regularization function for a range of $\epsilon$ values to find the the optimal regularization parameter at this distance, as shown in  in Figure \ref{fig:err_vs_eps_all}. The near-boundary optimal regularization parameter at the distance $d = h+R$ is the one that minimizes the percent error between the simulations and the theory. 

 \begin{figure}[ht]
    \centering
    \includegraphics[width=.75\textwidth]{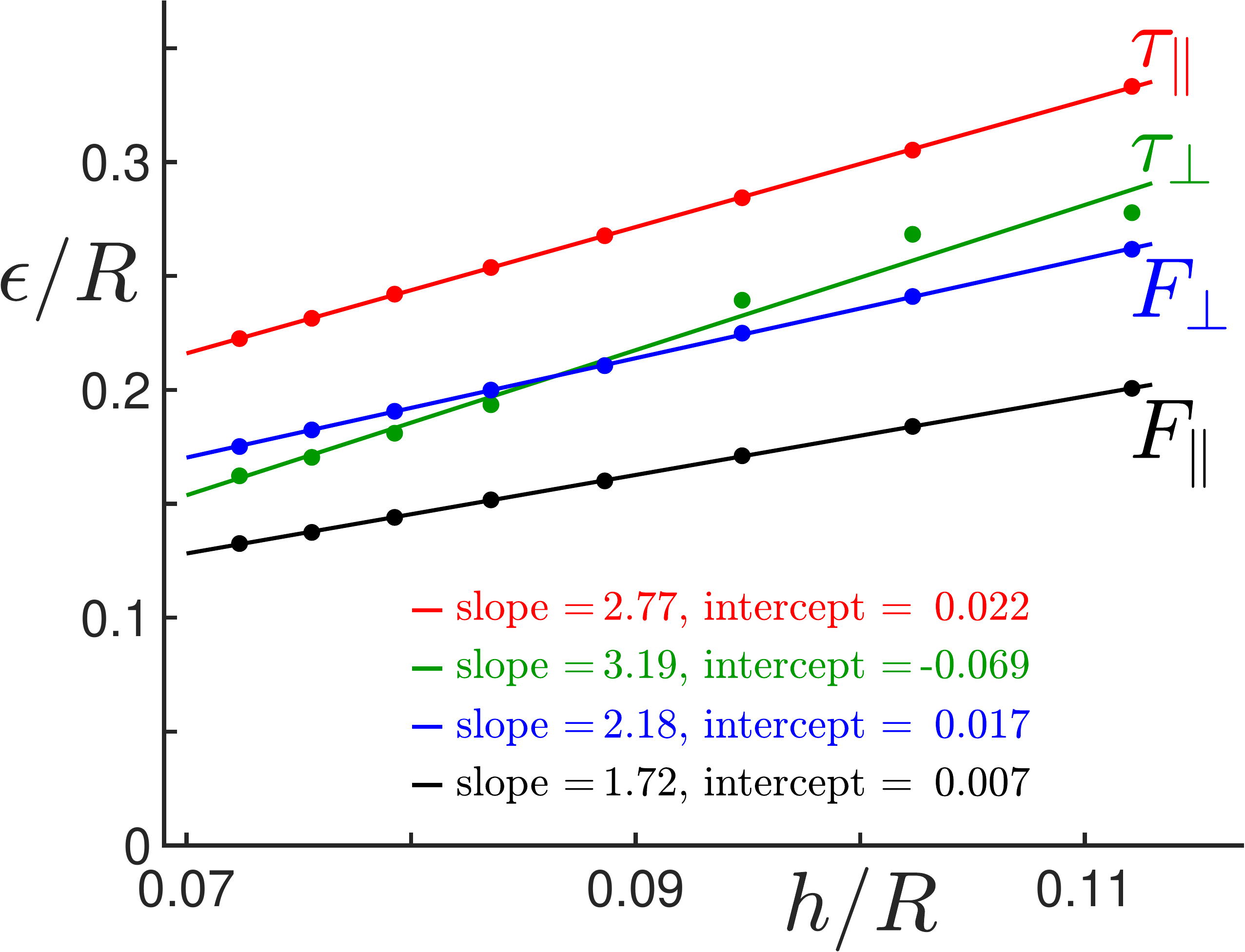}
\caption{Optimal regularization parameter $\epsilon$ plotted as a function of discretization length $h$. In these data, the edge of the sphere is located at a distance $h$ from the boundary. Optimal regularization parameters are found for each of the four motion types by identifying the minima of the curves shown in Figure \ref{fig:err_vs_eps_all}. The solid curves are linear fits to the data with coefficients shown. The data are fit very well except in the case of perpendicular translation $F_\perp$, which shows some scatter from the linear fit for the three lowest resolutions: $N=$ 1000, 1200, and 1400. The data show that there is no single optimal choice for the regularization parameter once the sphere is near the boundary. Whereas in free space, the optimal value of the regularization parameter is more similar for translation and rotation, as shown in Figure \ref{fig:fitting_optimal_blob_sizes}.} 
    \label{fig:epsilon_vs_h_fits}
\end{figure}

 The presence of the boundary breaks the symmetry of motion in free space and results in four different optimal regularization parameters rather than two very similar values of $\epsilon_{\rm free}$, as found for translation and rotation, see Figure \ref{fig:fitting_optimal_blob_sizes}(c)-(d). In free space, the values of $\epsilon_{\rm free}$ for translation (0.270) and rotation (0.248) differ by about 9\% for \SE\ with $N=1000$, whereas near the boundary the optimal values of $\epsilon$ vary by more than 50\% for the different motions. Figure \ref{fig:err_vs_eps_all} also shows that the varying $N$  changes the size of the optimal regularization parameter because the minima   occur at different locations $h$. Thus, not only does the average discretization size $h$ provide the threshold for the minimum gap between the edge of the sphere and the wall, but also the optimal regularization parameter depends on $h$ near the boundary.

Assuming $h$ is small, i.e., the sphere is moving near the boundary, we can Taylor expand the regularization parameter for each fundamental motion to first order,
\begin{equation}
    \epsilon(h) \approx \epsilon(0) + \left.\frac{d\epsilon}{dh}\right|_{h=0}h
    \label{eq:TaylorExpand_eps}
\end{equation}

In Figure \ref{fig:epsilon_vs_h_fits}, we show the results of optimizing $\epsilon$ at a boundary distance equal to the  discretization length $h$ over a range of values and fitting Eq. \ref{eq:TaylorExpand_eps} to the data. Near the boundary, small changes in $\epsilon$ lead to large changes in the percent error, while far from the boundary, the percent error is less sensitive to changes in $\epsilon$ (Figure \ref{fig:optimal_blob_sizes}). Note that the log scale in the vertical makes the rate of change in percent error as $\epsilon$ varies appear to be very rapid, but the slope of the curves in Figure \ref{fig:optimal_blob_sizes} near the minimum is typically 10, whereas in Figure \ref{fig:err_vs_eps_all} the slopes are of order $10^{-7}$ or smaller. 

Therefore, when a particular simulation requires some motion to be near a boundary, an alternative approach to reduce error is to determine the closest distance desired {\it a priori} and then to optimize the regularization parameter at that distance; motion further from the surface using same regularization parameter value will also have small errors. For the motion of a sphere, the fit values shown in Figure \ref{fig:epsilon_vs_h_fits} may be used to determine appropriate regularization parameters, though each component of motion would have to be simulated separately with a different value of $\epsilon$. This requirement makes our free-space optimization method a better choice in most situations.

\section{Numerical implementation of Lee and Leal theory \label{app:Lee_Leal}} 

The bipolar coordinate theory by Lee and Leal (1980) \cite{Lee_Leal_1980sphere_motion} gives a series solution to the force and torque on a sphere moving near an infinite boundary and includes a slip coefficient $\lambda$ that represents the degree of slip at the boundary from no-slip $\lambda=\infty$ to free-slip $\lambda=0$. We solve their equations listed below using both MATLAB and PYTHON GUI programs for convenience.
 The MATLAB code includes a GUI interface using MATLAB's AppDesigner functionality tested on R2021 and later. We also provide a PYTHON GUI program that contains the same basic features tested on PYTHON 3.12.1, which requires the PySimpleGUI, numpy, and pyperclip packages. The inputs to the GUI's are:
 \BE
 \item Radius of the sphere: the default is unity since all values are scaled using the radius.
 \item Viscosity ratio:  the range is $0> \lambda \leq \infty$. The default is \verb|"inf"|, i.e. infinity, which is equivalent to a no slip boundary near the sphere. Implementing $\lambda = 0$ was only included for perpendicular torque calculations since Lee and Leal developed an analytic expression. Otherwise, a very small nonzero value should be used in this limit.
 \item Number of terms: this is the truncation value for the infinite sums.
 \item Boundary distances: range $1<l=d/R\leq \infty$. These are boundary distances at which the theoretical values will be calculated. Dimensional values will be scaled by the sphere radius. When $l/R=1$, the theory diverges.
 \item Motion type: there are four radio buttons provided for selecting which of the four fundamental types of motion are to be calculated
 \EE
 
 The MATLAB GUI calls one of three primary functions, for which we have also coded PYTHON equivalents:
 \BE
 \item  \verb|motion_parallel.m| and \verb|motion_parallel.py|: calculate the force and torque for rotation or translation parallel to a plane interface.The inputs are: $l$, $\lambda$, $N$ and \verb|motion|, where \verb|motion=2| is for translation and \verb|motion=4| is for rotation following the notation in Lee and Leal.
 \item \verb|rotation_perpendicular.m| and \verb|rotation_perpendicular.py|: calculate the torque for rotation perpendicular to the boundary. The inputs are: $l$, $\lambda$, and $N$. Here all force components are zero.
 \item  \verb|translation_perpendicular.m| and  \verb|translation_perpendicular.py|: calculate the additional drag force for motion perpendicular to a plane interface. The inputs are: $l$, $\lambda$, and $N$. Here all torques are zero.
 \EE
 
The outputs from the function are the increase in the vector force and torque components so conversion to dimensional units is done by multiplying all force values by the Stokes drag $6\pi\mu Rv$ and by multiplying all torque values by the torque on a sphere far from a boundary $8\pi \mu R^3 \Omega$, where $\mu$ is the viscosity of the fluid, $R$ the radius of the sphere, $v$ is the speed of the sphere, and $\Omega$ is the angular frequency. 

Below are the system of equations developed by Lee and Leal that are solved in our MATLAB and PYTHON codes \cite{Lee_Leal_1980sphere_motion}. We have preserved their numbering for easy reference and included comments in the MATLAB and PYTHON codes that identify how each equation is used. 

In their work, translation perpendicular to a plane interface or rotation with axis perpendicular to the interface, result in $m=0$, whereas parallel motion results in $m=1$, in the equations below.

The equations result in seven unknown coefficients: $A_n^m$, $B_n^m$, $C_n^m$, $E_n^m$, $F_n^m$, $G_n^m$, and $H_n^m$. They note that $D_n^m=\hat{D}_n^m= \hat{C}_n^m=0$ because of equations (32) and (37) below. There are seven  coefficients for the other fluid denoted as $\hat{A}^m_n$, $\hat{B}^m_n$, $\hat{D}^m_n$, $\hat{E}^m_n$, $\hat{F}^m_n$,  $\hat{G}^m_n$, and $\hat{H}^m_n$.

\renewcommand{\theequation}{\arabic{equation}}
\begin{eqnarray*}
l\equiv d/R \qquad \qquad(2)\\
\eta_0 = -\cosh^{-1}(l) \qquad \text{see pg. 205}\\
\end{eqnarray*}
\setcounter{equation}{31}

for the upper fluid and all $n$ and $m$:
\begin{eqnarray}
\hat{A}^m_n =-\hat{B}^m_n,\ \hat{C}^m_n =-\hat{D}^m_n,\ \hat{E}^m_n =-\hat{F}^m_n,\ \text{ and }\ \hat{G}^m_n =-\hat{H}^m_n 
\end{eqnarray}

for $m=0$
\BSub
\begin{eqnarray}
-\frac{1}{2}nA^0_{n-1}+\frac{5}{2}A^0_n+\frac{1}{2}(n+1)A^0_{n+1} 
-nD^0_{n-1}+(2n+1)D^0_n-(n+1)D^0_{n+1} \nonumber\\
+n(n-1)E^0_{n-1}-2n(n+1)E^0_n +(n+1)(n+2)E^0_{n+1}=0 \nonumber\\ %\qquad (33a)
\end{eqnarray}
\ESub
and
\BSub
\begin{eqnarray}
-\frac{1}{2}nB^0_{n-1}+\frac{5}{2}B^0_n+\frac{1}{2}(n+1)B^0_{n+1} 
-nC^0_{n-1}+(2n+1)C^0_n-(n+1)C^0_{n+1} \nonumber\\
+n(n-1)F^0_{n-1}-2n(n+1)F^0_n 
+(n+1)(n+2)F^0_{n+1}=0 \nonumber\\%\qquad (34a)
\end{eqnarray}
\ESub

\setcounter{equation}{32}
while for $ m \geqslant 1,$:
\BSub
\stepcounter{equation}

\begin{eqnarray}
-\frac{1}{2}(n-m)A^m_{n-1}+\frac{5}{2}A^m_n+\frac{1}{2}(n+m+1)A^m_{n+1}
-(n-m)D^m_{n-1}+(2n+1)D^m_n\nonumber\\
-(n+m+1)D^m_{n+1} +\frac{1}{2}(n-m)(n-m-1)E^m_{n-1} -(n-m)(n+m+1)E^m_n \nonumber\\
+\frac{1}{2}(n+m+1)(n+m+2)E^m_{n+1}-\frac{1}{2}G^m_{n-1} 
+G^m_n-\frac{1}{2}G^m_{n+1} =0 \nonumber\\%(33b)
\end{eqnarray}
\ESub
and
\BSub
\stepcounter{equation}
\begin{eqnarray}
-\frac{1}{2}(n-m)B^m_{n-1}+\frac{5}{2}B^m_n+ 1/2(n+m+1)B^m_{n+1} 
-(n-m)C^m_{n-1}+(2n+1)C^m_n \nonumber\\
-(n+m+1)C^m_{n+1} +\frac{1}{2}(n-m)(n-m-1)F^m_{n-1} -(n-m)(n+m+1)F^m_n \nonumber\\
+\frac{1}{2}(n+m+1)(n+m+2)F^m_{n+1} 
-\frac{1}{2}H^m_{n-1} +H^m_n-\frac{1}{2}H^m_{n+1} = 0\nonumber\\  %(34b)
\end{eqnarray}
\ESub

\setcounter{equation}{36}
\begin{eqnarray}
D^m_n=\hat{D}^m_n=0 \qquad \text{for all} \quad n,m.
\end{eqnarray}

\setcounter{equation}{39}

for all $m$, $\eta$ and $\xi$, we require:
\begin{eqnarray}
-\frac{(n-m-1)}{2n-1}(F^m_{n-1}-\hat{F}^m_{n-1})+(F^m_n-\hat{F}^m_n)
-\frac{n+m+2}{2n+3}(F^m_{n+1}-\hat{F}^m_{n+1})\nonumber\\
=-\frac{1}{2}\left [\frac{1}{2n-1}(B^m_{n-1}-\hat{B}^m_{n-1})-\frac{1}{2n+3}(B^m_{n+1}-\hat{B}^m_{n+1}) \right ]\nonumber\\% \qquad (40)
\end{eqnarray}

\BSub
\begin{eqnarray}
\text{for }m=0: \quad H^0_n=\hat{H}^0_n %\qquad (41a)
\end{eqnarray}

for  $m \geqslant 1$: 
\begin{eqnarray}
-\frac{(n-m+1)}{2n-1}(H^m_{n-1}-\hat{H}^m_{n-1})+(H^m_n-\hat{H}^m_n)
-\frac{n+m}{2n+3}(H^m_{n+1}-\hat{H}^m_{n+1})\nonumber\\
=-\frac{1}{2}\left [-\frac{(n-m)(n-m+1)}{2n-1}(B^m_{n-1}-\hat{B}^m_{n-1})+\frac{(n+m)(n+m+1)}{2n+3}(B^m_{n+1}-\hat{B}^m_{n+1})\right ]\nonumber\\
%\qquad (41b)
\end{eqnarray}
\ESub

\setcounter{equation}{44}
for $m=0$:
\BSub
\begin{eqnarray}
G^0_n = \lambda\hat{G}^0_n %\qquad (45a)
\end{eqnarray}
\ESub

for all $m$:
\setcounter{equation}{53}

\begin{eqnarray}
E^m_n\sinh{(n+1/2)}\eta_0 + F^m_n\cosh{(n+1/2)}\eta_0
=X^m_n(\eta_0)-\frac{1}{(2n+3)\sinh{\eta_0}}[-Z^m_{n+1}(\eta_0) \nonumber\\
+C^m_{n+1}\sinh{(n+3/2)}\eta_0]+\frac{1}{(2n-1)\sinh{\eta_0}}[-Z^m_{n-1}(\eta_0)\nonumber\\
+C^m_{n-1}\sinh{(n-1/2)}\eta_0] \nonumber\\ %\qquad (54)
\end{eqnarray}

for $m=0$:
\BSub
\begin{eqnarray}
G^0_n\sinh{(n+1/2)}\eta_0+H^0_n\cosh{(n+1/2)}\eta_0 
=Y^0_n(\eta_0) %\qquad (55a)
\end{eqnarray}

for  $m \geqslant 1$: 
\begin{eqnarray}
G^m_n\sinh{(n+1/2)}\eta_0 + H^m_n\cosh{(n+1/2)}\eta_0\nonumber\\
=Y^m_n(\eta_0)+\frac{(n+m)(n+m+1)}{(2n+3)\sinh{\eta_0}}[-Z^m_{n+1}(\eta_0)
+C^m_{n+1}\sinh{(n+3/2)}\eta_0]\nonumber\\
-\frac{(n-m)(n-m+1)}{(2n-1)\sinh{\eta_0}}[-Z^m_{n-1}(\eta_0)
+C^m_{n-1}\sinh{(n-1/2)}\eta_0] %\quad \text{for } m \geqslant 1; %\qquad (55b)
\end{eqnarray}
\ESub

for all $m$:
\begin{eqnarray}
A^m_n\sinh{(n+1/2)}\eta_0+B^m_n\cosh{(n+1/2)}\eta_0
=\frac{-2}{\sinh{\eta_0}}[\frac{(n-m)}{2n-1}\{Z^m_{n-1}(\eta_0)\nonumber\\
-C^m_{n-1}\sinh{(n-1/2)}\eta_0\}
-\cosh{\eta_0}\{Z^m_n(\eta_0)-C^m_n\sinh{(n+1/2)\eta_0}\}\nonumber\\
+\frac{n+m+1}{2n+3}\{Z^m_{n+1}(\eta_0)
-C^m_{n+1}\sinh{(n+3/2)\eta_0}\}] \nonumber\\%\quad \text{for all } m. %\qquad (56)
\end{eqnarray}
and
\begin{eqnarray}
A^m_n=\lambda\hat{A}^m_n, \quad E^m_n=\lambda\hat{E}^m_n \quad and \quad G^m_n=\lambda\hat{G}^m_n. %\qquad (57)
\end{eqnarray}

The equations are rewritten so that the LHS has only coefficients, and all other terms are on the RHS. The matrix is constructed using a for-loop, where each column is built as shown in Eq. \ref{eq:theory_matrix}. The first column ($n=m-1$) and last column ($n =N+1$) are removed after construction because the sums should range from $\sum_{n=m}^N$, where $N$ is a truncation value chosen for a desired accuracy. The result is a square, banded matrix of size $N\times N$. The rows and columns are constructed as:
\normalsize
\BQ
\begin{pmatrix}\label{eq:theory_matrix}

n-1& n& n+1 & \dots &N+2\\
A^m_{(n-1)}& \dots\\
&A^m_n& \dots\\
&&A^m_{(n+1)} & \dots\\
B^m_{(n-1)}& \dots\\
&B^m_n& \dots\\
&&B^m_{(n+1)} & \dots\\
\vdots
\end{pmatrix}
\EQ

The resulting matrix is inverted and multiplied times the RHS to solve for the coefficients. The force and/or torque for the four fundamental motions are calculated using the following equations:

\setcounter{equation}{70}
\BE
\item Force for perpendicular translation: 
    \begin{eqnarray}
      F_x = F_y = 0, \nonumber \\
      F_z = \frac{-\sqrt{2}}{3}\sinh{\eta_0}\sum_n[C_n^0-(n+1/2)(A_n^0-B_n^0)] \nonumber \\
    \end{eqnarray}
    \begin{eqnarray}
    \tau_x = \tau_y = \tau_z = 0, \nonumber \\
    \end{eqnarray}

 \setcounter{equation}{74}
\item Force and torque for parallel translation:
    \begin{eqnarray}
    F_y = F_z = 0, \nonumber \\
    F_x = \frac{-\sqrt{2}}{6}\sinh{\eta_0}\sum_n[G_n^1-H_n^1+n(n+1)(A_n^1-B_n^1)]  \nonumber \\
    \end{eqnarray}

\setcounter{equation}{75}
    \begin{eqnarray}
       \tau_x = \tau_z = 0, \nonumber \\
       \tau_y = \frac{\sinh^2{\eta_o}}{12\sqrt{2}}\sum_n[(2+e^{(2n+1)\eta_0})\{n(n+1)(-2C_n^1-A_n^1\coth{\eta_0}) \nonumber \\
       -(2n+1+\coth{\eta_0})G_n^1\}
       +(2-e^{(2n+1)\eta_0})\{n(n+1)B_n^1\coth{\eta_0} \nonumber \\
       +(2n+1+\coth{\eta_0})H_n^1\}].\nonumber\\
    \end{eqnarray}

  \item  Torque for perpendicular axis rotation
\setcounter{equation}{81}

    \begin{eqnarray}
        F_x=F_y=F_z=0
    \end{eqnarray}
    
    \begin{eqnarray}
        \tau_x = \tau_y = 0, \nonumber \\
        \tau_z = \frac{\sinh^2{\eta_0}}{\sqrt{2}}\sum_nn(n+1)(-G_n^0+H_n^0).
    \end{eqnarray}

    \item Force and torque for parallel axis rotation
\setcounter{equation}{85}
    \begin{eqnarray}
        F_x = \frac{-\sqrt{2}}{6}\sinh{\eta_0}\sum_n[G_n^1-H_n^1+n(n+1)(A_n^1-B_n^1)], \nonumber \\ 
        F_y = F_z = 0 \nonumber \\
    \end{eqnarray}

    \begin{eqnarray}
        \tau_x = \tau_z = 0, \nonumber \\ 
        \tau_y = -\frac{1}{3}-\frac{\sinh^2{\eta_0}}{12\sqrt{2}}\sum_n[(2+e^{(2n+1)\eta_0})\nonumber \\ 
        \times\{n(n+1)(-2C_n^1-A_n^1\coth{\eta_0}) \nonumber\\ 
        -(2n+1+\coth{\eta_0})G_n^1\} \nonumber\\ 
        +(2-e^{(2n+1)\eta_0})\{n(n+1)B_n^1\coth{\eta_0} 
        +(2n+1+\coth{\eta_0})H_n^1\}]. %\nonumber%\qquad (87) 
    \end{eqnarray}

\EE
The MATLAB and PYTHON codes can solve the series solutions quickly even when $10^4$ terms are included using a computer with 16GB of RAM. This number of terms is only necessary for very small boundary distances and changes the result in 5-6 digit, whereas for many cases $N=100$ is sufficient. 

We compared our results from Lee and Leal \cite{Lee_Leal_1980sphere_motion}  with the values from our calculations using $N=500$ terms and found that the maximal difference is in the sixth digit of accuracy and is generally in the seventh or eighth digit. Our code does not implement $\lambda = 0$, so the values used $\lambda = 10^{-8}$. At this level of precision, it is unclear whose calculations are correct, but the results show that the numerical implementation matches the work of Lee and Leal to a high degree of accuracy. We note that they appear to have a mistake in the value for $\lambda = 10$ and $l=1.6$, which we believe should be $\tau = - 5.26852\times 10^{-5}$ but is reported as being $\tau = - 5.26852\times 10^{-3}$. The former value matches our results very well, as do all of the other values in the table, so we believe the value in the table is incorrect. The supplementary materials include a MATLAB script (Lee\_and\_Leal\_tabulated\_values.m) that calculates these differences, which are available here: \url{https://drive.google.com/drive/folders/1w-U42eVbh17HiOrBwUJB6SW0ttMy6n0s?usp=share_link}.

\end{document}